\documentclass[12pt]{article}
\usepackage{amsmath,amsfonts,amssymb,bm,slashed,bbm}
\usepackage{graphicx,color}\usepackage{float}

\usepackage[x11names]{xcolor}

\usepackage{listings}  \newcommand{\code}[1]{\lstinline|#1|}      

\newcommand{\eq}[2]{\begin{align}\label{#1}#2\end{align}}

\newcommand{\nn}{\nonumber}

\newcommand{\pa}{\partial}

\newcommand{\Eq}[1]{\eqref{#1}}
\renewcommand{\Eq}[1]{\eqref{#1}}

\newcommand{\al}{\alpha}\newcommand{\be}{\beta}
\newcommand{\ga}{\gamma}\newcommand{\Ga}{\Gamma}
\newcommand{\la}{\lambda}
\newcommand{\om}{\omega}

\newcommand{\x}{\vec x}

\begin{document}

\title{Zeta function in Coulomb background}
\author{
		M. Bordag\footnote{bordag@jinr.ru}\\
		\small     Bogoljub Laboratory of Theoretical Physics, Joint Institute for Nuclear Research\\
		\small 141980 Dubna, Russian Federation\\
		\small Institute for Theoretical Physics, University Leipzig,
		IPF 231101, D-04081 Leipzig, Germany
		\and
		I.G. Pirozhenko\footnote{pirozhen@theor.jinr.ru} \\
		\small     Bogoljub Laboratory of Theoretical Physics, Joint Institute for Nuclear Research\\
		\small 141980 Dubna, Russian Federation\\
}

\maketitle 
\begin{abstract}
We calculate the one-loop vacuum energy of a massive scalar field confined to the interior of a sphere with Dirichlet boundary conditions and subject to a Coulomb background potential $\al/r$. The spectrum is determined by the zeros of a confluent hypergeometric (Kummer) function, and the corresponding zeta function is constructed and analytically continued to $s=-1/2$. The renormalization is performed by subtracting the divergent part of the zeta function, defined through the heat kernel expansion, which removes the non-decreasing powers of the mass and leaves a finite, physically normalized vacuum energy. While the heat kernel coefficients for the pure Bessel (Coulomb-free) case are well known, the Coulomb contributions to the coefficients are not; we determine them by matching the analytic continuation of the zeta function to the general form of the divergent part, obtaining terms proportional to $\al$ and $\al^2$. A key technical point is that the uniform asymptotic expansion of the mode generating function cannot be used in full; a part must be retained unexpanded, and an integral representation of the digamma function is used to carry out the analytic continuation. In the infinite-volume limit the renormalized vacuum energy tends to a constant $E_0^{ren,\infty}(\al,m)=\al e(\al/m)$, with a dimensionless function $e$ that depends only on the ratio $\al/m$. This limit shows that a finite, physically meaningful renormalized vacuum energy can be defined for a massive scalar field in the background of a Coulomb source. For $m\to\infty$ it vanishes in accordance with the normalization condition, while for $m\to0$ it diverges, so that the present calculation does not apply to a massless field, pointing to the need of another vacuum state in that case.
\end{abstract}
\section{\label{T1}Introduction}

Vacuum polarization in a Coulomb background has been a subject of interest since the early days of quantum electrodynamics (QED). One of the classical results is the calculation of the vacuum-polarization correction to the Coulomb potential, which leads to the well-known Uehling potential \cite{wich56-101-843}. Such corrections are of particular importance for atomic physics and, more generally, for the interpretation of high-precision measurements of atomic energy levels. The Coulomb field therefore provides one of the simplest and at the same time most important backgrounds in which quantum-field-theoretical effects can be studied explicitly. Interest also comes from critical fields, as they are conjectured to appear in superheavy nuclei or heavy ion collisions. A great deal of work was done in the 1980s and collected, for instance, in the book \cite{greiner85}. This topic is of current theoretical and experimental interest, see the recent paper \cite{gras22-106-013003} and citations therein. However, the Coulomb field is long-range, which put its own subtlest and some standard methods, especially in scattering theory, do not apply.

Closely related to vacuum polarization are the vacuum energy and the induced vacuum charge in a Coulomb background field. The corresponding local densities have been calculated by several different methods. Their ultraviolet renormalization is closely related to charge renormalization and leads, in the case of the vacuum-polarization correction to the Coulomb potential, to counterterms resulting from the Uehling potential. For a pure Coulomb background, nearly explicit expressions are known for the Green function and the heat kernel, in particular by exploiting the relation of the Coulomb problem to the four-dimensional harmonic oscillator, \cite{duru79-84-185}, or to the O(2,1) algebra, \cite{mils82-90-447}. All these make the Coulomb background an especially useful example for studying the interplay between spectral properties, ultraviolet divergences, and renormalization.

The local vacuum-energy density, however, cannot in general be integrated directly to obtain the complete vacuum energy, because of its singular behavior at the position of the external source. This is not an exceptional feature of the Coulomb problem. A closely related phenomenon occurs in the Casimir effect: the local energy density diverges near a material boundary, although the total Casimir energy can nevertheless be defined consistently by spectral or other global methods. For instance, the Casimir energy between parallel plates can be calculated by several independent approaches, and the corresponding force is not only well defined but has also been measured with high precision. For a comprehensive discussion of the Casimir effect and its various formulations, see, for instance, the book \cite{BKMM}.
Another, more severe problem, specific to the Coulomb background, arises from the logarithmic (Coulomb) phase, which does not allow one to integrate the local energy density at $r\to\infty$ unless the contributions from particle and antiparticle are added together. This problem was observed long ago and a possible solution was indicated in \cite{doll66-45-801}.

 In addition, the most powerful method of the   regularization, which is necessary anyway, namely the zeta functional regularization,  hits a specific problem. In the sum over the spectrum, the bound states accumulate at the mass threshold and do not decrease for any power $s$. Hence, the zeta function for the Coulomb potential in infinite space may not exist, except for a neutral field with repulsive potential, as we consider here. For a charged field in full QED, where bound states exist for each sign of the charge, there is in addition the question about the definition of the vacuum state, which should account for the soft photons (infraparticles).

These observations motivate the problem considered in the present paper. We study the one-loop vacuum energy of a scalar field confined to the interior of a finite sphere and subject to a Coulomb (or Newtonian) background centered at the origin, as an external-potential spectral problem. As the simplest realization of this problem, we consider a neutral scalar field obeying the equation
\eq{i1}{ \left(-\Delta+\frac{\al}{r}+m^2\right)\phi(\x) &=
	\om^2\phi(\x),&\phi(\x){|_{|\x|=R}}&=0,
}
and impose a Dirichlet boundary condition at the radius $r=R$. This resembles the non-relativistic equation for hydrogen or a particle in the field of a point mass. But, at this stage, we do not attach a specific physical interpretation to the coupling $\al$. In this equation, assuming natural units, $\al$ has the dimension of an inverse length.

For $\al=0$, the corresponding problem for the electromagnetic field was studied by Boyer \cite{boye69-10-1729}, in the context of the electromagnetic Casimir energy of a perfectly conducting spherical shell. His result ruled out Casimir's electron model. Subsequent calculations considered scalar fields as well as the interior and exterior regions of the sphere separately. A comprehensive treatment of these problems, including their spectral and zeta-functional formulations, can be found in the book \cite{kirsten01}.

The Coulomb background is also technically interesting from the point of view of the heat kernel and its asymptotic expansion. Knowledge of the heat kernel coefficients is equivalent to knowledge of the ultraviolet divergences and, consequently, of the counterterms required for their renormalization. For smooth backgrounds, the standard iterative procedures for constructing the heat kernel, related to the Schwinger--DeWitt expansion, provide a systematic way of obtaining these coefficients. For manifolds with boundaries, a large number of results are known as well; see, for example, \cite{vass03-388-279}. The Coulomb potential presents a somewhat different situation. It is not sufficiently singular at the origin to prevent the coefficients $a_1$ and $a_2$ from being integrated at $r=0$, but, in the terminology relevant here, it is a slowly decreasing long-range potential. For such backgrounds, the global heat kernel coefficients in the presence of a boundary do not appear to have been determined explicitly. For infinite volume, the reason is the slow decrease of the local coefficients after taking the coincidence limit (but see \cite{slob96-11-1729}). For finite volume, this problem has not been considered. An intriguing question is what happens when the infinite-volume limit is taken afterwards.

It is interesting to mention that in one dimension the situation is different. With a Coulomb background there is no problem at infinity, but a severe problem at the origin. The corresponding Laplace operator may not be self-adjoint and, even after a self-adjoint extension, logarithmic terms may occur in the heat kernel expansion, as recently investigated in \cite{falo20-53-465402}.

To be specific, we calculate the zeta function
\eq{i2}{ \zeta(s) &=\sum_{l=0}^\infty\sum_{n=1}^\infty (2l+1)\om_{l,n}^{-2s}
}
and the related vacuum energy
\eq{i3}{ E_0(s) &= \frac{\hbar}{2}\mu^{2s}\zeta(s-\frac12)
}
for the spectrum of the one-particle energies $\om_{l,n}=\sqrt{\la_{l,n}^2+m^2}$ with the eigenvalues $\la_{l,n}^2$
of the radial equation
\eq{i4}{
	\left(-\frac{\pa^2}{\pa r^2}-\frac{2}{r}\frac{\pa}{\pa r}
	+\frac{l(l+1)}{r^2}+\frac{\al}{r}\right)\phi_{l,n}(r)
	&= \la_{l,n}^2 \phi_{l,n}(r), & \phi_{l,n}(R)&=0.
}
In the following we put the parameter $\mu $ , which describes the arbitrariness which comes in with the regularization, to unity, $\mu=1$, since this arbitrariness will be removed by our renormalization scheme, see below.

The above setting provides one of the simplest settings in which the Casimir effect can be studied in the presence of a long-range potential in a finite volume. As we shall see, however, the problem is technically considerably more involved than its formulation might suggest. The eigenfunctions $\phi_{l,n}(r)$ are known explicitly in terms of confluent hypergeometric (Kummer) functions,
\eq{i5}{ \phi_{l,n}(r) &= r^l e^{i \la_{l,n}r}
	M\left(l+1+\frac{\al}{-2i \la_{l,n}}
	,2l+2,-2i \la_{l,n}r\right).
}

The expression \Eq{i3} provides a representation of the vacuum energy in zeta-functional regularization. It is ultraviolet finite for $s>\frac32$ and has to be analytically continued to $s=0$. For this purpose, we apply the method developed in \cite{bord97-56-4896}. First, we transform the radial sum in \Eq{i2} into a contour integral. For this purpose we introduce the {\it mode generating function}
\eq{i6}{\Phi_l(\la)&=
	e^{i\la R}M\left(l+1-\frac{\al}{i2\la },2l+2,-2i\la R\right),
}
whose zeros give the eigenvalues $\la_{l,n}$. The function $\Phi_l(\la)$ is meromorphic in the upper half of the complex $\la$-plane and has a smooth continuation to the real axis. We then apply the Cauchy theorem and deform the integration contour towards the imaginary axis, $\la\to i\xi$. This gives the representation
\eq{i7}{ \zeta(s) &= \frac{\sin(\pi s)}{\pi} 
	\sum\limits_{l=0}^\infty (2l+1)
	\int\limits_m^\infty d\xi \ (\xi^2-m^2)^{-s} \pa_\xi \ln\Phi_l(i\xi)
}
with
\eq{i7a}{\Phi_l(i\xi) &= e^{- \xi R}
	M\left(l+1+\frac{\al}{2\xi},2l+2,2\xi R\right).
}
The last expression  for $\zeta(s)$ is real as long as the background potential is not critical, i.e., as long as its possible discrete levels remain below the mass threshold $m$. In that case possible bound states do not show up explicitly in \Eq{i7}. The sum and the integral converge for $s>\frac32$ and define a meromorphic function of $s$, with simple poles in the region  $\Re s\le\frac32$. Technically, the analytic continuation is performed by subtracting and adding a suitable asymptotic expression $\ln\Phi_l^{as}(i\xi)$. The details of this procedure are given in the next section.

The renormalization of the vacuum energy can be formulated in terms of the zeta function together with the heat kernel expansion. This procedure has been presented repeatedly in the literature, for example in \cite{BKMM}, and we therefore give here only the basic steps. Representing the powers of the frequencies in \Eq{i2} by a proper-time integral and interchanging the sum and the integral leads to the corresponding heat kernel,
\eq{i8}{K(t) &= \sum\limits_{l=0}^\infty (2l+1)
	\sum_{n=1}^\infty e^{-t( \la_{l,n}^2+m^2)},
}
and to the representation
\eq{i9}{ \zeta(s)&= \frac{1}{\Gamma(s)}\int_0^\infty \ dt\,t^{s-1}K(t)
}
for the zeta function.

For small $t$, the heat kernel has the asymptotic expansion
\eq{i10}{K(t) &\simeq \frac{1}{(4\pi t)^{3/2}}\sum_{n\ge0}a_nt^n e^{-tm^2},
}
where the {\it heat kernel coefficients} $a_n$ have integer indices for smooth background fields and acquire half-integer indices in the presence of boundaries. Inserting the expansion \Eq{i10} into the zeta function and retaining the coefficients with $n\le2$, we define
\eq{i11}{
	\nn\zeta^{div}(s) &=
	-\frac{m^4 a_0}{32\pi^2}\left[\frac{1}{s+1/2}-\frac{1+\ln(m/2)}{2}\right] 
	-\frac{m^3 a_{1/2}}{12\pi^{3/2}}
	+\frac{m^2a_1}{16\pi^2} \left[\frac{1}{s+1/2}-\left(1+2\ln(m/2)\right)\right]
\\  &~~~
	+\frac{ma_{3/2}}{8\pi^{3/2}}
	-\frac{a_2}{16\pi^2}\left[\frac{1}{s+1/2}-2\left(1+\ln(m/2)\right)\right].
}
as the divergent part of the zeta function. It contains the pole at $s=-1/2$ together with a particular finite part. We use $\zeta^{div}(s)$ to define the renormalized zeta function, $\zeta^{ren}$, by
\eq{i12}{\zeta^{ren} &= \lim_{s\to-1/2}
	\left(\zeta(s)-\zeta^{div}(s)\right).
}
Since the heat kernel expansion \Eq{i10} is a semiclassical asymptotic expansion in inverse powers of the mass and, at the same time, an expansion in powers of $\hbar$, the subtraction of $\zeta^{div}(s)$ in \Eq{i12} removes the non-decreasing powers of the mass while retaining the genuine quantum contributions. In this sense, the subtraction removes the classical contributions and can be regarded as a physically motivated normalization condition. It also provides a unique prescription for the renormalized zeta function and, consequently, for the vacuum energy. This normalization condition is applicable to massive models in general and can also be applied to massless models whenever the latter can be defined as the massless limit of a massive theory, as recently observed in \cite{bord26-8-79}. 

A comment is in order here regarding the arbitrary dimensional parameter $\mu$, which is commonly introduced together with the zeta-functional regularization in equations \Eq{i2} or \Eq{i3}. It disappears, first formally, in the limit $s\to0$ (in \Eq{i3}), and later, after renormalization, once the normalization is fixed by requiring the renormalized vacuum energy to be unique. This is especially the case when subtracting all contributions that do not decrease at large mass, to which $\ln\mu$ would belong.

The above procedure requires knowledge of the heat kernel coefficients. For $\al=0$, these coefficients are well known; see, for instance, \cite{vass95-36-3174}. They were recently reconsidered in \cite{bord26-8-79}, and we display them in Appendix \ref{Ahk}. In this case, the eigenvalues are determined by the zeros of a Bessel function, $\la_{l,n}=j_{l,n}$. Instead, for the Coulomb background, the corresponding coefficients are not known explicitly, as already mentioned. However, they can be inferred from the analytic continuation of the zeta function to $s=0$ together with the knowledge of the existence of the heat kernel expansion. The latter, in finite volume, is guaranteed by a general theorem, \cite{bran90-15-245}. We carry out this analysis in Section \ref{T3}. Originally, this method was applied in \cite{bord96-37-895} to determine the boundary contributions to the heat kernel coefficients.

The main technical step in our procedure is the analytic continuation of $\zeta(s)$, \Eq{i2}, to $s=-1/2$. Since this is not possible directly there, nor in \Eq{i7}, we define a suitable $\Phi^{as}_l(\xi)$, such that with the definition
\eq{i13}{\zeta(s)^{as}&=
		 \frac{\sin(\pi s)}{\pi} 
			\sum\limits_{l=0}^\infty (2l+1)
			\int\limits_m^\infty d\xi \ (\xi^2-m^2)^{-s} \pa_\xi \ln\Phi_l^{as}(i\xi)
}
the limit
\eq{i14}{\zeta^{num} &\equiv\lim_{s\to -1/2}\left(\zeta(s)-\zeta(s)^{as}\right)
}
is finite. Here $\zeta(s)^{as}$ is defined in such a way that the difference appears under the sign of summation and integration,
\eq{i15}{\zeta^{num} & =
		 \frac{-1}{\pi}
	\sum\limits_{l=0}^\infty (2l+1)
	\int\limits_m^\infty d\xi \ (\xi^2-m^2)^{1/2}
	 \pa_\xi\left[ \ln\Phi_l(i\xi)- \ln\Phi_l^{as}(i\xi)\right],
}
and that these converge. This allows us to put $s=-1/2$. This part of the zeta function is called {\it numeric part}, since it can be calculated only numerically. It is a function of $\al$ and $m$ alone. After defining $\zeta(s)^{as}$, \Eq{i13}, we subtract it in \Eq{i14} from the zeta function and must add it back in \Eq{i12},
\eq{i16}{\zeta^{ren}&=\zeta^{num}+\lim_{s\to-1/2}
	\left(\zeta(s)^{as}-\zeta(s)^{div}\right).
}
We introduce the separate notation,
\eq{i17}{ \zeta^{an}&=
	\lim_{s\to-1/2}
	\left(\zeta(s)^{as}-\zeta(s)^{div}\right),
}
since $\zeta(s)^{an}$ will occupy the main part of the paper. In simpler examples, $\zeta(s)^{an}$ had nearly explicit, analytic expressions, hence the naming {\it analytic part}. In earlier examples it was the beginning of the uniform asymptotic expansion of the mode generating function on the imaginary axis for large orbital momentum and uniform in the radial momentum $\xi$ (this will be modified in the next section). The pole parts compensate in \Eq{i17} by construction. The analytic continuation in $\zeta(s)^{as}$ is the main technical content of this paper, and at the same time the most difficult one, with a number of surprising features.

The paper is organized as follows. In the next section we calculate the asymptotic part of the zeta function. This is the main part of the paper. Then follows a section where we derive the heat kernel coefficients. In the fourth section we calculate the analytical and numerical parts of the zeta function. In the last section we discuss the properties of the vacuum energy. At the end come the appendices.

Throughout most of the paper, until the last section, we put the radius $R=1$. The dependence on $R$ can be restored by the substitutions
\eq{i18}{m&\to mR,& \al &\to \al R, & E_0&\to \frac{E_0}{R}.
}
The zeta function is dimensionless. Most of the paper is written in terms of the zeta function. The final, renormalized vacuum energy is given by the formula
\eq{i18a}{ E_0^{ren}&= \frac{1}{R}{\zeta^{ren}}_{|_{\al\to\al R,m\to mR}}
}
in terms of $\zeta^{ren}$, \Eq{i16}.

The orbital momentum $l$ enters in most places through 
\eq{i19}{  \nu\equiv l+\frac12,
}
which we will use throughout the paper. Aslo we use the notation $\ga$ for the Euler constant and $\ga_1(x)$ for the first Stieltjes constant.
\section{\label{T2}The asymptotic part of the Zeta function}
We defined the asymptotic part of the zeta function in \Eq{i13}, which followed from \Eq{i7} by the substitution of the mode generating function by an asymptotic approximation.
For later use we note this expression after substitution $\xi= \nu z$,
\eq{2.2}{ \zeta^{as}(s) &=2 \frac{\sin(\pi s)}{\pi}\sum_\nu \nu\int_{m/\nu}^\infty\, dz\,
	 \left[(\nu z)^2-m^2\right]^{-s}\pa_z\ln\Phi^{as}_l(i\nu z)
}
We define the asymptotic approximation by
\eq{2.3}{ \ln\Phi^{as}_l(i\nu z)\ &= \sum_{k=-1}^3
	 \nu^{-k}h_k(z) + \nu^{-5}h_5(z)   +w(\nu,z).
}
The functions $h_k(z)$ and $w(\nu,z)$ are derived in Appendix \ref{A}.

This choice is inspired by the corresponding expansion in the case without Coulomb potential ('Bessel' case), which was first done in \cite{bord96-37-895} and \cite{bord97-56-4896}. It amounts to the asymptotic expansion of $ \ln\Phi^{as}_l(\nu z)$, in that case a Bessel function, for large $\nu$ which is uniform in $z$. In our case \Eq{i6} with a Kummer function (we call it 'Kummer' case), it needs a modification which will be explained below. Before that, however, a more general remark is in order. The choice of $ \ln\Phi^{as}_l(i\nu z)$ is not restricted in the first place since we subtract it and add it back as explained in the preceding section whenever restricted by the convergence requirements. Another, more technical restriction follows from the aim to construct the analytic continuation in $s$. In the initial expression with the Kummer function this is clearly impossible. If one were to construct the asymptotic expansion directly from the Temme coefficients in Appendix \ref{B}, this would be very complicated. So we found that the partial expansion \Eq{2.3} keeps the necessary convergence properties and at the same time allows for a feasible analytic continuation. Here the difficult point is the argument in the gamma function, see eq. \Eq{2.5} below, which has $\nu$ both in a numerator and in a denominator. Of course, the origin of this contribution is the well known logarithmic Coulomb phase.

The functions $h_k(z)$ are parts of the uniform asymptotic expansion in $\nu$ and have contributions from several powers of the coupling $\al$,
\eq{2.4}{ h_k( z) &= \sum_{p=0}^3 h_{p,k}(z)\al^p,
	\qquad (k=1,2,3,5),
}
where $p=0$ is the Bessel case, and a non-expanded contribution
\eq{2.5}{  w(\nu,z) &= \ln\Gamma(2\nu+1)
	-\ln\Ga\left(\nu+\frac12+\frac{\al}{2\nu z}\right).
}
These functions are shown in Appendix \ref{A}. The reason behind this choice is the following. When expanding the complete expression, i.e., including $w(\nu,z)$, for large $\nu$, we get in $\zeta^{num}$, \Eq{i15}, a difference $\ln \Phi_l(i\xi)-\ln\Phi^{as}_l(i\xi)$ which decreases for large $\nu$, but not for large $z$, and the corresponding $z$-integration would not converge.

With these splits, we rewrite the zeta function \Eq{2.2},
\eq{2.6}{  \zeta^{as}(s) &=  \sum_{k=1}^3  Z_k+Z_5+Z_w ,
}
according to the sum in \Eq{2.3}, with
\eq{2.7}{ Z_k &=2\frac{\sin(\pi s)}{\pi}
	\sum_{\nu} \nu^{1-k}	\int_{m/\nu}^\infty dz\,
	\left[(\nu z)^2-m^2\right]^{-s}\pa_z h_k(z),
	\quad k=1,2,3,5,
\\\nn 	&\equiv   	\sum_{p=0}^3 Z_{p,k}\, \al^p
	\\\nn 	 Z_w &=2\frac{\sin(\pi s)}{\pi}
\sum_{\nu} \nu	\int_{m/\nu}^\infty dz\,
		\left[(\nu z)^2-m^2\right]^{-s}\pa_z \ w(\nu,z)
}
and
\eq{2.8}{        Z_{p,k} &= 2\frac{\sin(\pi s)}{\pi}
	\sum_{\nu} \nu^{1-k}	\int_{m/\nu}^\infty dz\,
	\left[(\nu z)^2-m^2\right]^{-s}\pa_z h_{p,k}(z).
}
In the following these expressions will be calculated separately for each $p$ and $k$, and separately for the Kummer and Bessel parts,
\eq{2.8a}{ Z_{p,k} &\equiv  Z^K_{p,k}(\nu)  + Z^B_{p,k}(\nu)
}
Here, in \Eq{2.6}, expressions appear which do not decrease in $\nu$ sufficiently fast to make the sums over $\nu$ convergent in the individual expressions \Eq{2.8}. We call these 'bad' terms; but in section \ref{T25} we will see that the bad terms compensate when added together.

According to \Eq{2.8a}, we split also the zeta function, \Eq{2.6},
\eq{2.8b}{ \zeta^{as}(s) &= \zeta_K^{as}(s) + \zeta_B^{as}(s).
}
For $\zeta_B^{as}(s)$ we use the corresponding results from \cite{bord26-8-79}. The calculation of the $Z^K_{p,k}$ in \Eq{2.8a}, entering \Eq{2.7} and \Eq{2.6}, is the topic of the following subsections.

\subsection{\label{T21} The part $Z_{1,1}$ }

In this and in the following subsections we calculate $Z_{p,k}^K$ and use simplified notations
\eq{2.1.1.0}{ Z_{p,k}^K \to Z_{p,k}.
}
Inserting $ h^K_{1,1}(z)$ from \Eq{A1.7} into \Eq{2.8}, we get as starting point
\eq{2.1.1}{  Z_{1,1} &=2 \frac{\sin(\pi s)}{\pi}\al
	\sum_{l=0}^\infty \int_{m/\nu}^\infty dz \ ((\nu z)^2-m^2)^{-s}
	\pa_z \frac{\ln (z+\sqrt{1+z^2})+\ln\nu}{2z}.
}
By the substitution $z\to z/\nu$ it turns into
\eq{2.1.2}{  Z_{1,1} &= 2\frac{\sin(\pi s)}{\pi}\al
	\sum_{l=0}^\infty      \int_{m}^\infty dz \ ( z^2-m^2)^{-s}
	\pa_z \frac{\nu}{2z}\ln (z+\sqrt{\nu^2+z^2}).
}
Expanding the expression under the derivative for $\nu\to\infty$,
\eq{2.1.3}{  \frac{\nu}{2z}\ln (z+\sqrt{\nu^2+z^2}) &=
	 h_{1,1}^{as}+O\left(\frac{1}{\nu^2}\right), \quad\mbox{with}\quad
	 h_{1,1}^{as} = \frac12+\frac{\nu\ln\nu}{2z}
}
we split the expression \Eq{2.1.2} into two parts,
\eq{2.1.3a}{ Z_{1,1} &= Z_{1,1}^{as} +  Z_{1,1}^{reg},
}
where
\eq{2.1.4}{  Z_{1,1}^{as} &=2\frac{\sin(\pi s)}{\pi}\al
	 \sum_{l=0}^\infty      \int_{m}^\infty dz \ ( z^2-m^2)^{-s}
	\pa_z  h_{1,1}^{as},
\\\nn 	 Z_{1,1}^{reg} &=2\frac{\sin(\pi s)}{\pi}\al
\sum_{l=0}^\infty      \int_{m}^\infty dz \ ( z^2-m^2)^{-s}
\pa_z  \left( \frac{\nu}{2z}\ln (z+\sqrt{\nu^2+z^2}) -h_{1,1}^{as}\right).
}
In $Z_{1,1}^{as}$, the z-integration gives
\eq{2.1.5}{  Z_{1,1}^{as} &=-\al
	\sum_{l=0}^\infty 
	\frac{\nu  m^{-2 s-1} \log (\nu ) \sin (\pi  s) \Gamma (1-s) \Gamma \left(s+\frac{1}{2}\right)}{\pi ^{3/2}}.
}
The sum over $l$ does not converge and we keep it unevaluated. This is the first 'bad' term,
\eq{2.1.5a}{ Z_{1,1}^{'bad'} &=  Z_{1,1}^{as}.
}

In $ Z_{1,1}^{reg}$ we interchange sum and integration,
\eq{2.1.6}{  Z_{1,1}^{reg} &= 2\al\frac{\sin(\pi s)}{\pi}
	 \int_{m }^\infty dz \ ( z^2-m^2)^{-s} 	\pa_z  S_1
}
with
\eq{2.1.7}{ S_1 &= \sum_{l=0}^\infty
	\left(\frac{\nu}{2z}\ln (z+\sqrt{\nu^2+z^2})
	 - \frac12-\frac{\nu\ln\nu}{2z}\right).
}
This sum converges, but cannot be done explicitly. We need its behavior for large $z$. Therefore we apply the Abel-Plana formula (see \Eq{A2.1}) and get two contributions accordingly,
\eq{2.1.8}{ S_1=S_1^A+S_1^B.
}
Similarly we split,
\eq{2.1.10a}{  Z_{1,1}^{reg} &= Z_{1,1}^{reg,A}+Z_{1,1}^{reg,B},
}
with \eq{2.1.8a}{ Z_{1,1}^{reg,A}=
	 2\al\frac{\sin(\pi s)}{\pi}
	\int_{m }^\infty dz \ ( z^2-m^2)^{-s} 	\pa_z  S_1^A,
\\\nn
	 Z_{1,1}^{reg,B}=
	2\al\frac{\sin(\pi s)}{\pi}
	\int_{m }^\infty dz \ ( z^2-m^2)^{-s} 	\pa_z  S_1^B.
}
In the first integral we carry out the $\nu$-integration,
\eq{2.1.9b}{ S_1^A &= \int_0^\infty d\nu \
	 \left(\frac{\nu}{2z}\ln (z+\sqrt{\nu^2+z^2})
	- \frac12-\frac{\nu\ln\nu}{2z}\right)
	=-\frac{z}{4}.
}
The remaining $z$-integration gives
\eq{2.1.9}{ Z_{1,1}^{reg,A} &=
2\al	\frac{sin(\pi s)}{\pi}\int_m^\infty dz \ (z^2-m^2)^{-s}\pa_z S_1^A
\\\nn 	&=\frac{-\al m^{1-2s}}{8\sqrt{\pi}}\Ga(1-s)\Ga(-1/2+s),
\\\nn 	&= -\frac{\al m^2}{8 \pi  \left(s+\frac{1}{2}\right)}
	+\frac{\al m^2 (2 \log (\frac{m}{2})+1)}{8 \pi}
	+O\left(\left(s+\frac{1}{2}\right)^1\right).
}
In the last line we continued to $s=-1/2$ and obtained a pole and a regular term,
\eq{2.1.9a}{ Z_{1,1}^{reg,A,pp}&=-\frac{\al m^2}{8\pi},
\\\nn  Z_{1,1}^{reg,A,reg}&=
	-\frac{\al m^2 (-2 \log (m)-1+\log (4))}{8 \pi }.
}

In the second integral, $Z_{1,1}^{reg,B}$, we turn the variable, $\nu\to i\nu$, and get a contribution to $g(\nu)$ (see \Eq{A2.1}) only from $\nu\le z$,
\eq{2.1.10b}{ g(\nu)&=  \frac{\nu}{z}\ln\left(\frac{z}{\nu}
	+\sqrt{\left(\frac{z}{\nu}\right)^2-1}\right),\quad (\nu<z),
}
resulting in
\eq{2.1.10}{ S_1^B &= \frac{1}{z}\int_0^z \frac{d\nu\,\nu}{1+\exp(2\pi\nu)}
	\ln\left(\frac{z}{\nu}+\sqrt{\left(\frac{z}{\nu}\right)^2-1}\right),
}
which obviously converges.

Inserting $S_1^B$ into $Z_{1,1}^{reg}$, \Eq{2.1.10a}, we get a z-integration which converges for $\frac12<s<1$ and we are left with the task of the analytic continuation to $s=-1/2$.
For this,  we split
\eq{2.1.12}{ S_1^{B} &= S_1^{Bas}+S_1^{Breg}
}
by adding and subtracting the first term of the expansion for $z\to\infty$,
\eq{2.1.13}{ \ln\left(\frac{z}{\nu}+\sqrt{\left(\frac{z}{\nu}\right)^2-1}\right)
	 &= \ln\frac{2z}{\nu}+\dots\,.
}
We get
\eq{2.1.13a}{ S_1^{Bas} &=
	\frac{1}{z}\int_0^z \frac{d\nu\,\nu}{1+\exp(2\pi\nu)}
	\ln\frac{2z}{\nu},
\\\nn
S_1^{Breg} &=
	\frac{1}{z}\int_0^z \frac{d\nu\,\nu}{1+\exp(2\pi\nu)}
	\left[
	\ln\left(\frac{z}{\nu}+\sqrt{\left(\frac{z}{\nu}\right)^2-1}\right) -\ln\frac{2z}{\nu}              \right].
}
In $ S_1^{Bas}$, the $z$-integration can be done explicitly if we extend it to the whole axis, thus redefining $ S_1^{Bas}$ and $ S_1^{Breg}$ by
\eq{2.1.13b}{ S_1^{Bas} &=
	\frac{\gamma\,\pi^2 + \pi^2\ln(8\pi z)
		- 3\,\gamma_1^{(0,1)}(1,\frac{1}{2})  
		+ 3\, \gamma_1^{(0,1)}(1,1)  }{48\pi^2 z},
\\\nn
	S_1^{Breg} &=
	-\frac{1}{z}\int_z^\infty \frac{d\nu\,\nu}{1+\exp(2\pi\nu)}
	\ln\frac{2z}{\nu}
	+
	\frac{1}{z}\int_0^z \frac{d\nu\,\nu}{1+\exp(2\pi\nu)}
	\left[
	\ln\left(\frac{z}{\nu}+\sqrt{\left(\frac{z}{\nu}\right)^2-1}\right) -\ln\frac{2z}{\nu}              \right],
}
where we used the notation $\ga_1(x)$ for the first Stieltjes constant.
Calculating the integral in $ S_1^{Bas}$ numerically allows us to find the numerical values,
\eq{2.1.13c}{
\ga_1^{(0,1)}\left(1,\frac{1}{2}\right)= 6.68364,\
\ga_1^{(0,1)}(1,1)=
	0.707386.
}
Now $S_1^{Breg}$ decreases like $z^{-2}$ for $z\to\infty$ which allows us to put $s=-1/2$ in the $z$-integration in \Eq{2.1.8a} directly.

According to the subdivision \Eq{2.1.12}, we divide
\eq{2.1.14}{ Z_{1,1}^{reg,B} &=  Z_{1,1}^{reg,B,as}+Z_{1,1}^{reg,B,reg}.
}
First, we calculate the contribution from $S_1^{B,as}$. The integration can be done explicitly and delivers
\eq{2.1.16} { Z_{1,1}^{reg,B,as} &=
	\frac{(-1)^s m^{-2 s-1} \sin (\pi  s) }{48 \pi ^{5/2} \Gamma \left(\frac{1}{2}-s\right) \Gamma (s)}        
\Bigg[-2 \pi  \csc (\pi  s) \left(-3
		\ga_1^{(0,1)}\left(1,\frac{1}{2}\right)+3 \ga_1^{(0,1)}(1,1)
\right.\\\nn&~~~\left.		
		+\pi ^2 \log (m)+\pi
		^2 (-1+\gamma +\log (8 \pi ))\right)
	\\\nn&~~~+i \Gamma \left(\frac{1}{2}-s\right) \Gamma \left(s+\frac{1}{2}\right)
\\\nn&		\left(-6 \ga_1^{(0,1)}\left(1,\frac{1}{2}\right)+6 \ga_1^{(0,1)}(1,1)+2 \pi ^2
		\log (m)-\pi ^2 \psi ^{(0)}\left(s+\frac{1}{2}\right)-i \pi ^3+\gamma  \pi ^2
\right.\\\nn&~~~\left.+2 \pi ^2 \log (\pi )+\pi ^2 \log
		(16)\right)
		+\pi ^2 \Gamma (1-s) \Gamma (s) \left(\psi ^{(0)}\left(\frac{1}{2}-s\right)+\gamma -2+\log
		(4)\right)\Bigg] .    
}
Expanding around $s=-1/2$, we see a double pole, a single pole and a regular part,
\eq{2.1.16a}{
 Z_{1,1}^{reg,B,as,ppp} &= \frac{\al}{96\pi},
 \\\nn
 Z_{1,1}^{reg,B,as,pp} &=
 	\frac{\al \left(-3 \ga_1^{(0,1)}\left(1,\frac{1}{2}\right)+3
 		\ga_1^{(0,1)}(1,1)+\pi ^2 (-1+\gamma +\log (8)+\log (\pi ))\right)}{48 \pi ^3},
\\\nn
Z_{1,1}^{reg,B,as,reg} &=
	\frac{\al}{576 \pi ^3}
\\\nn & \times\Bigg[	-24 \log (m) \left(-3 \ga_1^{(0,1)}\left(1,\frac{1}{2}\right)+3
		\ga_1^{(0,1)}(1,1)+\pi ^2 \log (4 \pi )\right)
\\\nn&
		-72 \ga_1^{(0,1)}(1,1)-12 (\log (64)-6)
		\ga_1^{(0,1)}\left(1,\frac{1}{2}\right)
\\\nn &
		+36 \log (4) \ga_1^{(0,1)}(1,1)-12 \pi ^2 \log
		^2(m)-24 \gamma  \pi ^2 \log (2 m)
\\\nn&
		+\pi ^2 \left(-2 \pi ^2+3 \log (4) (5 \log (4)-12)+12 \gamma  (\log (16)-2)+12 (\log
		(4)-2) \log (\pi )\right)\Bigg].
}

The second part in \Eq{2.1.14} is
\eq{2.1.17}{  Z_{1,1}^{reg,B,reg} &= \frac{-2\al}{\pi}
	\int_m^\infty dz\,\sqrt{z^2-m^2}\pa_z S_1^{Breg}
}
with $S_1^{Breg}$ defined in \Eq{2.1.13a}.
This is a double integration. It is convergent and we can put $s=-1/2$. This integration is left for numerical evaluation. For convenience, to get rid of the derivative, one may integrate by parts,
\eq{2.1.18}{  Z_{1,1}^{reg,B,reg} &= \frac{2\al}{\pi}
	\int_m^\infty dz\,\frac{1}{\sqrt{z^2-m^2}}\,
\\\nn &~~~~~ \times\left\{
	\int_0^z \frac{d\nu\ \nu}{1+\exp(2\pi \nu)}
	\ln\left[\frac12\left(1+\sqrt{1-\left(\frac{\nu}{z}
		\right)}\right)\right]
+	\int_z^\infty \frac{d\nu\ \nu}{1+\exp(2\pi \nu)}
		\ln\frac{2z}{\nu}
		\right\}	,
}
with no surface terms.

\subsubsection{\label{T211}The pole parts and the regular terms}
Here we collect the results of the calculation done so far.
For $s\to -1/2$, there are contributions from
$ Z_{1,1}^{reg,A}$, \Eq{2.1.9}, from
$Z_{1,1}^{reg,B,as}$, \Eq{2.1.16a}, and
$ Z_{1,1}^{reg,B,reg}$, \Eq{2.1.18},
\eq{2.1.19}{Z_{1,1} &= Z_{1,1}^{reg,A} +Z_{1,1}^{reg,B,as}+Z_{1,1}^{reg,B,reg},
}
which have  pole terms and regular terms,
\eq{2.1.20}{ Z_{1,1} &=
	 \frac{	Z_{1,1}^{ppp}}{ \left(s+\frac{1}{2}\right)^2}
	+ \frac{	Z_{1,1}^{pp}}{ \left(s+\frac{1}{2}\right)}
	+	Z_{1,1}^{reg}+O\left(s+\frac12\right).
}
These are
\eq{2.1.21}{
Z_{1,1}^{ppp} &=\frac{\al}{96\pi},
\\\nn
Z_{1,1}^{pp} &= \frac{\al \left(-3 \ga_1^{(0,1)}\left(1,\frac{1}{2}\right)+3
	\ga_1^{(0,1)}(1,1)+\pi ^2 (-1+\gamma +\log (8)+\log (\pi ))\right)}{48 \pi ^3}-\frac{\al m^2}{8
	\pi },
\\\nn
Z_{1,1}^{reg} &=  Z_{1,1}^{reg,A,reg}+Z_{1,1}^{reg,B,as,reg}+Z_{1,1}^{reg,B,reg},
}
with $Z_{1,1}^{reg,A,reg}$, \Eq{2.1.9a}, $Z_{1,1}^{reg,B,as,reg}$, \Eq{2.1.16a}, and $Z_{1,1}^{reg,B,reg}$, \Eq{2.1.18}.
All contributions are explicit up to $Z_{1,1}^{reg,B,reg}$, which must be calculated numerically.

Finally we collect the contributions from  $m=0$  and from  $m\to\infty$.
For $m=0$ we return to $Z_{1,1}^{reg}$, \Eq{2.1.4}. In that case the integration over $z$ is explicit and the subsequent sum over $\nu$ as well, resulting in
\eq{2.1.22}{ Z_{1,1}^{m=0} &=
	\frac{2 \al \left(4^s-1\right) s \zeta (2 s) \sin (\pi  s) \Gamma (-s) \Gamma \left(s+\frac{3}{2}\right)}{\pi
		^{3/2} (2 s+1)^2}.
}
For $m\to\infty$,  the non-decreasing part of $Z_{1,1}^{reg}$ comes from $Z_{1,1}^{reg,A,reg}$, \Eq{2.1.9a}, and $Z_{1,1}^{reg,B,as,reg}$, \Eq{2.1.16a}, and reads
  \eq{2.1.23}{ Z_{1,1}^{inf} &=Z_{1,1}^{reg,A,reg}+Z_{1,1}^{reg,B,as,reg}
 \\\nn
 &=
 \frac{\al}{576 \pi ^3}
 \\\nn & ~~\times\Bigg[ -12 \left(2 \log (m) \left(-3 \ga_1^{(0,1)}\left(1,\frac{1}{2}\right)+3
 	\ga_1^{(0,1)}(1,1)+\pi ^2 \left(\log (4 \pi )-6 m^2\right)\right)
 \right.\\\nn&~~~~\left.
 	-3 \log (4)
 	\ga_1^{(0,1)}(1,1)+\pi ^2 \log ^2(m)+\pi ^2 (2 \gamma  \log (2 m)-(\log (4)-2) \log (\pi ))\right)
 \\\nn&	~~~~
 	-72
 	\ga_1^{(0,1)}(1,1)-12 (\log (64)-6) \ga_1^{(0,1)}\left(1,\frac{1}{2}\right)
 \\\nn&	~~~~
 	+\pi ^2
 	\left(-72 m^2 (\log (4)-1)-2 \pi ^2+12 \gamma  (\log (16)-2)+3 \log (4) (5 \log (4)-12)\right)\Bigg],
}
where, again, $\ga_1$ is defined in \Eq{2.1.13c}.
\subsection{\label{T2b1} The part $Z_{1,2}$}

Inserting $h^K_{1,2}(z)$ from \Eq{A1.7} into \Eq{2.8}, we get as starting point
\eq{2.2b.1}{ Z_{1,2} &= 2\al\frac{\sin(\pi s)}{\pi}\sum_{l=0}^\infty   \frac{1}{\nu}   \int_{m/\nu}^\infty dz \ ( (\nu z)^2-m^2)^{-s}
	\pa_z \frac{-1}{4\sqrt{1+z^2}}.
}
We separate the sum and carry out the $z$-integration. Defining
\eq{2.2b.2}{ w_{12} &=
	2\al\frac{\sin(\pi s)}{\pi}   \frac{1}{\nu}   \int_{m/\nu}^\infty dz \ ( (\nu z)^2-m^2)^{-s}
	\pa_z \frac{-1}{4\sqrt{1+z^2}},
\\\nn & =
	\frac{1}{2} \al \nu  s \left(m^2+\nu ^2\right)^{-s-1},
}
we represent
\eq{2.2b.3}{  Z_{1,2} &= \sum_{l=0}^\infty    w_{12} .
}
We need the analytic continuation in $s$. For this, we split the summand,
\eq{2.2b.4}{  w_{12}&=w_{12}^{as}+w_{12}^{reg}
}
with
\eq{2.2b.5}{ w_{12}^{as} &=\frac{\al\, s}{2}\nu^{-1-2s},
	&w_{12}^{reg} &=\frac{\al}{4}\left(1-\frac{1}{\sqrt{1+m^2/\nu^2}}\right).
}
In $w_{1,2}^{reg}$ we have put $s=1/2$ as we need it only in \Eq{2.2b.9} below.
Accordingly, we divide
\eq{2.2b.6}{ Z_{1,2} &=  Z_{1,2}^{as}  +  Z_{1,2}^{reg}.
}

In $ Z_{1,2}^{as} $, the sum over $l$ is explicit and results in
\eq{2.2b.7}{  Z_{1,2}^{as} &= \frac{1}{2} \al \left(2^{2 s+1}-1\right) s \,
	 \zeta (2 s+1).
}
The expansion in $s$ gives
\eq{2.2b.8}{ Z_{1,2}^{as} =O\left(s+\frac12\right)
}
and from here  no pole contribution and no regular contribution follow.

In the regular part, $Z_{1,2}^{reg}$, the summation converges and we can put $s=1/2$ directly,
\eq{2.2b.9}{  Z_{1,2}^{reg} &=
	 \sum_{l=0}^\infty     w_{12}^{reg}.
}
This is a convergent sum and may be used for numerical evaluation. Because of \Eq{2.2b.8}, we have no pole contribution and the regular contribution comes only from \Eq{2.2b.9},
\eq{2.2b.10}{  Z_{1,2} &=  Z_{1,2}^{reg}.
}

Below we need also the behavior for large $m$. For this we apply the Abel-Plana formula, \Eq{A2.1} to \Eq{2.2b.9}. Part A is a simple integration and delivers
\eq{2.2b.11}{  Z_{1,2} ^A &= \frac{\al}{4}m^{-s},
}
and,  especially for $s=0$,
\eq{2.2b.12}{   Z_{1,2} ^A &= \frac{\al}{4}m.
}
In part B we get
\eq{2.2b.13}{ g(\nu) &=  -\frac{\al}{2}\frac{\nu}{\sqrt{m^2-\nu^2}},~~~(\nu<m),
}
and
\eq{2.2b.14}{  Z_{1,2} ^B  &= -\frac{\al}{2}
	\int_0^m \frac{d\nu}{1+\exp(2\pi\nu)}  \frac{\nu}{\sqrt{m^2-\nu^2}}.
}
From here it is obvious that the B-part decreases for large $m$ and the leading asymptotic part comes only from the A-part,
\eq{2.2b.15}{  Z_{1,2}^{inf} &=
	\frac{\al}{4}m-\frac{\al}{96m}
	+O\left(\frac{1}{m^3}\right)
	~~~\mbox{for}~~m\to\infty.
}
 For $m=0$ we return to \Eq{2.2b.3}. In that case the sum can be done and results in
  \eq{2.2b.16}{ Z_{1,2}^{m=0} &= \frac{\al}{2} \left(2^{2 s+1}-1\right)
 	 s\, \zeta (2 s+1).
 }

\subsection{\label{T22}The part $Z_{1,3}$}

Here we insert $h^K_{1,3}(z)$  from \Eq{A1.7},
\eq{2.2.1}{h^K_{1,3}(z) &=
	\frac{-1}{48\sqrt{1+z^2}}-\frac{1}{8(1+z^2)^{3/2}}
	+\frac{5}{16(1+z^2)^{5/2}} +\frac{1}{48z}
}
into \Eq{2.8}. We start with taking the last term from this expression and insert it  into \Eq{2.8},
\eq{2.2.2}{ Z_{1,3}^{last} &= 2\al\frac{\sin(\pi s)}{\pi}\sum_{l=0}^\infty
	   \frac{1}{\nu^2}   \int_{m/\nu}^\infty dz \ ( (\nu z)^2-m^2)^{-s}
	\pa_z \frac{1}{48z}.
}
The integration over $z$ can be done immediately, resulting in
\eq{2.2.2a}{ Z_{1,3}^{last} &= \sum_{l=0}^\infty
\frac{-\al m^{-2 s-1} \sin (\pi  s) \Gamma (1-s) \Gamma \left(s+\frac{1}{2}\right)}{48 \pi ^{3/2} \nu }.
}
The sum over $l$ does not converge and it is not regularized by $s$. We keep it unevaluated. This is the second 'bad' term and we denote it as
\eq{2.2.3}{ Z_{1,3}^{'bad"} &=  Z_{1,3}^{last}.
}

Now we consider the first 3 terms in \Eq{2.2.1}. When inserted into \Eq{2.8}, the $z$-integration can be done and results in
\eq{2.2.4}{ w_{1,3} &=
	2\al\frac{\sin(\pi s)}{\pi}\frac{1}{\nu^2}\int_{m/\nu}^\infty dz \
	((\nu z)^2-m^2)^{-s}\pa_z  \left(h^K_{1,3}(z)-\frac{1}{48z}\right),
\\\nn & =\frac{\al \sin (\pi  s) \Gamma (1-s) \Gamma \left(s+\frac{1}{2}\right) \left(m^2+\nu ^2\right)^{-s-\frac{5}{2}} \left(m^4+2
	m^2 \nu ^2 (8 s+5)-2 \nu ^4 (2 s (5 s+6)+3)\right)}{24 \pi ^{3/2} \nu }.
}
With it, $Z_{1,3}$ appears as sum,
\eq{2.2.5}{Z_{1,3} &= \sum_{l=0}^\infty w_{1,3}.
}
Further, we split $w_{1,3} $ into parts,
\eq{2.2.6}{ w_{1,3} &=c_{13} \sum_{k=1}^3   p_k (\nu^2+m^2)^{k-s-7/2},
}
with
\eq{2.2.7}{
	c_{13} &= \frac{\al \sin (\pi  s) \Gamma (1-s) \Gamma \left(s+\frac{1}{2}\right)}{24 \pi ^{3/2} \nu},
\\\nn
	p_1 &=
	-20 m^4 s^2-40 m^4 s-15 m^4,
\\\nn
	p_2&=40 m^2 s^2+64 m^2 s+22 m^2,
\\\nn
	p_3&=20 s^2-24 s-6.
}  
Now, for $k=1$ and $k=2$, the summation over $l$ is convergent. We split accordingly
\eq{2.2.8}{w_{1,3} &=w_{1,3}^a+w_{1,3}^b.
}
with
\eq{2.2.9}{
	w_{1,3}^a&=c_{1,3}\sum_{k=1}^2 p_k(\nu^2+m^2)^{k-s-7/2},
 & w_{1,3}^b&= c_{1,3}p_3(\nu^2+m^2)^{-s-1/2}.
}
Due to the convergence, we put $s=-1/2$ and get
\eq{2.2.10}{& \sum_{l=0}^\infty w_{1,3}^a
\\\nn&=
\frac{\al}{48 \pi }
\bigg[-2 \psi ^{(0)}\left(\frac{1}{2}-i m\right)-2 \psi ^{(0)}\left(i m+\frac{1}{2}\right)
+5 i m \psi
	^{(1)}\left(\frac{1}{2}-i m\right)
	-5 i m \psi ^{(1)}\left(i m+\frac{1}{2}\right)
\\& ~~~~~~~~~~~~~~
	+20 \gamma +5 \log (256)
	+24 \psi
	^{(0)}\left(\frac{1}{2}\right)\bigg].
}
For $k=3$, the sum over $l$ is logarithmically divergent at $s=-1/2$. We split,
\eq{2.2.11}{\nu^{-1}(\nu^2+m^2)^{-s-1/2} &=
	\nu^{-2s-2}+\nu^{-2s-2}\left(\left(1+\frac{m^2}{\nu^2}\right)^{-s-1/2}-1\right),
}
and accordingly
\eq{2.2.12}{w_{1,3}^b   = w_{1,3}^{b1}+w_{1,3}^{b2}.
}

Again, the $l$-sum of the first part can be done analytically,
\eq{2.2.13}{\sum_{l=0}^\infty w_{1,3}^{b1} &=
	-\frac{\al\left(4^{s+1}-1\right) \left(10 s^2+12 s+3\right) \zeta (2 s+2) \sin (\pi  s) \Gamma (1-s) \Gamma
		\left(s+\frac{1}{2}\right)}{12 \pi ^{3/2}}.
}
In $s=-1/2$, it has a double pole, a simple pole and a regular part,
\eq{2.2.14}{Z_{1,3}^{ppp} &= -\frac{\al}{96 \pi },
\\\nn Z_{1,3}^{pp} &=-\frac{\al (-3+\gamma +\log (8))}{48 \pi },
\\\nn Z_{1,3}^{reg,a} &=
\frac{\al \left(24 \left(\gamma _1+3+\log (512)\right)+\pi ^2-84 \log ^2(2)-72 \gamma  (\log (2)-1)\right)}{576 \pi }.
}
The sum with $w_{1,3}^{b2}$ is convergent and we put $s=-1/2$,
\eq{2.2.15}{ {Z_{1,3}^{reg,b2}} &=
	\sum_{l=0}^\infty
	\frac{\al \log \left(\frac{m^2}{\nu ^2}+1\right)}{48 \pi  \nu }.
}
This sum is left for numerical evaluation. Now the regular part consists of two pieces,
\eq{2.2.16}{Z_{1,3}^{reg,b} &= Z_{1,3}^{reg,b1}+Z_{1,3}^{reg,b2},
}
and the complete regular part is
\eq{2.2.16a}{Z_{1,3}^{reg} &= Z_{1,3}^{reg,a}+Z_{1,3}^{reg,b}
}
with
\eq{2.2.16b}{Z_{1,3}^{reg} &=
	\frac{\al}{576 \pi }
	\Bigg[-24 \psi ^{(0)}\left(\frac{1}{2}-i m\right)-24 \psi ^{(0)}\left(i m+\frac{1}{2}\right)
\\\nn&~~	+60 i m \left(\psi
		^{(1)}\left(\frac{1}{2}-i m\right)
	-\psi ^{(1)}\left(i m+\frac{1}{2}\right)\right)
\\\nn&	~~	+24 \left(\gamma _1+3+\log (32)\right)+\pi ^2-12 \left(7
		\log ^2(2)+\gamma  (\log (64)-2)\right)\Bigg]
\\\nn& ~~   +\frac{\al}{48\pi}
	\sum_{l=0}^\infty
	\frac{  \log \left(\frac{m^2}{\nu ^2}+1\right)}{  \nu }.
}

For the asymptotic behavior at $m\to\infty$, we have logarithmic contributions and a constant term. The calculation was done applying the Abel-Plana formula to $Z_{1,3}^{reg2}$ and resulted in
\eq{2.2.17}{Z_{1,3}^{inf} &=
	\frac{\al \left(\log ^2(m)+2 (\gamma +\log (4)) \log (m)-2 \gamma _1+\frac{\pi ^2}{12}+2 \log ^2(2)+\gamma  \log (16)\right)}{48 \pi
	}.
}
For $m=0$, the initial sum in \Eq{2.2.5} can be done immediately and delivers
\eq{2.2.18}{Z_{1,3}^{m=0}&=
	-\frac{\al \left(4^{s+1}-1\right) \left(10 s^2+12 s+3\right) \zeta (2 s+2) \sin (\pi  s) \Gamma (1-s) \Gamma
		\left(s+\frac{1}{2}\right)}{12 \pi ^{3/2}}.
}

\subsection{\label{T23}The part $Z_{2,3}$}

Inserting $h^K_{2,3}(z)$ from \Eq{A1.7}, into \Eq{2.8}, we get as starting point
\eq{2.3.1}{ Z_{2,3} &= 2\al^2\frac{\sin(\pi s)}{\pi}
	\sum_{l=0}^\infty   \frac{1}{\nu^2}   \int_{m/\nu}^\infty dz \ ( (\nu z)^2-m^2)^{-s}
	\pa_z \frac{\sqrt{1+z^2}}{8z^2}.
}
After substitution $z\to z/\nu$ this formula turns into
\eq{2.3.2}{ Z_{2,3} &=2 \al^2\frac{\sin(\pi s)}{\pi}
	\sum_{l=0}^\infty
	 \int_{m}^\infty dz \ ( z^2-m^2)^{-s}
	\pa_z\tilde h(z).
}
with
\eq{2.3.3}{ \tilde h(z) &= \frac{\sqrt{\nu^2+z^2}}{8\nu z^2}.
}
To proceed, we define an asymptotic part,
\eq{2.3.4}{ h_{as}(z)&=\frac{1}{8z^2}
\left(1+\frac{z}{\nu}-\frac{z}{\nu+1}
	+\frac{z^2}{2(1+\nu^2)^2}\right),
}
and split
\eq{2.3.5}{ Z_{2,3} &=  Z_{2,3}^{as}+Z_{2,3}^{reg}
}
with
\eq{2.3.6}{ Z_{2,3}^{as} &= 2\al^2\frac{\sin(\pi s)}{\pi}
	 \sum_{l=0}^\infty
	\int_{m/\nu}^\infty dz \ ( z^2-m^2)^{-s}
	\pa_z h_{as}(z).
}
and
\eq{2.3.7}{ Z_{2,3}^{reg} &= 2\al^2\frac{\sin(\pi s)}{\pi}
	\sum_{l=0}^\infty
	\int_{m/\nu}^\infty dz \ ( z^2-m^2)^{-s}\pa_z
	\left(\tilde h(z)- h_{as}(z)\right).
}
This split is motivated by the possibility to take the $z$-integration in $ Z_{2,3}^{as} $ explicitly and to improve the convergence in $ Z_{2,3}^{reg}$.

\subsubsection{\label{T231}The part $Z_{2,3}^{as}$}

For $ Z_{2,3}^{as}$ we get after integration
\eq{2.3.8}{  Z_{2,3}^{as} &= 2\al^2\frac{\sin(\pi s)}{\pi}
		\sum_{l=0}^\infty
	\left[
	 -\frac{1}{8} \pi  s m^{-2
		s-2} \csc (\pi  s)
	 -\frac{m^{-2 s-1} \Gamma (1-s) \Gamma \left(s+\frac{1}{2}\right)}{8 \sqrt{\pi } \left(\nu ^2+\nu \right)}
	 \right]
	. 
}
The first term in the square bracket is just the third 'bad' term. We denote it by
\eq{2.3.9}{  Z_{2,3}^{bad} &=	\sum_{l=0}^\infty
\frac{-\al^2}{4}   s \,m^{-2 s-2}.
}
Now we denote the second term by $ Z_{2,3}^{as}$ and note
\eq{2.3.10}{  Z_{2,3}^{as} &=  2\al^2\frac{\sin(\pi s)}{\pi}
	\sum_{l=0}^\infty
	\frac{-m^{-2 s-1} \Gamma (1-s)
		\Gamma \left(s+\frac{1}{2}\right)}
		{8 \sqrt{\pi } \left(\nu ^2+\nu \right)} .
}
The sum can be taken delivering
\eq{2.3.10a}{  Z_{2,3}^{as} &=
	-\frac{\al^2 m^{-2 s-1} \sin (\pi  s) \Gamma (1-s) \Gamma \left(\frac{1}{2} (2 s+1)\right)}{2 \pi ^{3/2}}.
}
From here a pole part results,
\eq{2.3.10b}{  Z_{2,3}^{as,pp}&=-\frac{\al^2}{8 \pi },
}
and no regular part.

\subsubsection{\label{T232}The part $Z_{2,3}^{reg}$}
The treatment of the second part in \Eq{2.3.5} is more complicated. First we integrate over $z$,
\eq{2.3.11}{ Z_{2,3}^{reg} &= 2\al^2\frac{\sin(\pi s)}{\pi} \sum_{l=0}^\infty T(\nu)
}
with
\eq{2.3.12}{
&T(\nu) \equiv 	\int_m^\infty dz (z^2-m^2)^{-s}\pa_z
	\left( \tilde h(z)-h_{as}(z)\right),
\\\nn &
	=\frac{  \pi  s m^{-2 (s+1)}}{	\sin (\pi  s)}+
\frac{s \Gamma (-s-1) \Gamma \left(s+\frac{3}{2}\right) \left(\left(m^2+\nu ^2\right)^{-s-\frac{1}{2}}-m^{-2 s-1} \,
	_2F_1\left(-\frac{1}{2},s+\frac{3}{2};\frac{1}{2};-\frac{\nu ^2}{m^2}\right)\right)}{8 \sqrt{\pi } \nu }.
}
This expression is true, but somehow misleading since the first term has no $\nu$. This is compensated by the second term. Instead, we define
\eq{2.3.13}{
	 t_1&=\frac{s (s+2) \left(2 s^2+11 s+9\right) m^{-2 s} \Gamma (-s-2) \Gamma \left(s+\frac{3}{2}\right) \left(\frac{m}{\nu }\right)^{2 s}}{16
		\sqrt{\pi } \nu ^2},
\\\nn
	t_2&=-\frac{\nu  s \left(2 s^3+15 s^2+33 s+22\right) \Gamma (-s-2) \Gamma \left(s+\frac{3}{2}\right) \left(m^2+\nu ^2\right)^{-s-\frac{3}{2}}}{16
		\sqrt{\pi }},
\\\nn
	t_3&=\frac{-\frac{6 m \nu }{m^2+\nu ^2}+\frac{4 m}{\nu }-2 \tan ^{-1}\left(\frac{\nu }{m}\right)+\pi }{16 m},
}
with
\eq{2.3.14}{ T(\nu) &= t_1+t_2+t_3
}
and correspondingly,
\eq{2.3.14a}{  Z_{2,3}^{reg} &=
	 Z_{2,3}^{reg,1}+ Z_{2,3}^{reg,2}+ Z_{2,3}^{reg,3}.
}
These sums can be done. For the first one we arrive at
\eq{2.3.15}{
Z_{2,3}^{reg,1}&= 2\al^2\frac{\sin(\pi s)}{\pi}
	\sum_{l=0}^\infty t_1
\\\nn &=
  \frac{\al^2 \left(2^{2 s+2}-1\right) s \left(2 s^3+15 s^2+31 s+18\right) \zeta (2 (s+1)) \sin (\pi  s) \Gamma (-s-2) \Gamma
 	\left(\frac{1}{2} (2 s+3)\right)}{8 \pi ^{3/2}}.
}
For the second we define
\eq{2.3.16}{ t_2 &= \frac{\nu}{(\nu^2+m^2)^{s+3/2}}\tilde t_2
}
and split
\eq{2.3.17}{  \tilde t_2\equiv t_2^{as}+t_2^{reg}
}
with
\eq{2.3.18}{  t_2^{as} &= \nu^{-2-2s}\tilde t_2,
	&t_2^{reg} &=
	\left[\frac{\nu}{(\nu^2+m^2)^{s+3/2}}-\nu^{-2-2s}\right]
	\tilde t_2
}
and
\eq{2.3.18a}{  Z_{2,3}^{reg,2}=
	 Z_{2,3}^{reg,2,as}+ Z_{2,3}^{reg,2,reg}.
}

The summation in $ t_2^{as}$ is explicit and   that with $ t_2^{reg}$ converges at $s=-1/2$,
\eq{2.3.19}{ Z_{2,3}^{reg,2,as}&= 	\sum_{l=0}^\infty t_2^{as}
\\\nn&=-\frac{\al^2 \left(2^{2 s+2}-1\right) s \left(2 s^3+15 s^2+33 s+22\right) \zeta (2 (s+1)) \sin (\pi  s) \Gamma (-s-2) \Gamma
	\left(\frac{1}{2} (2 s+3)\right)}{8 \pi ^{3/2}}
\\\nn
 Z_{2,3}^{reg,2,reg}&= 	 \sum_{l=0}^\infty t_2^{reg}
=\frac{3 \al^2 \left(H_{-i m-\frac{1}{2}}+H_{i m-\frac{1}{2}}+\log (16)\right)}{8 \pi }
}
where $H$ is a {\it Harmonic number}.
The summation in $t_3$ can also be done analytically by differentiating the sum with respect to $m$, after which it can be done and integrating back,
\eq{2.3.20}{Z_{2,3}^{reg,3}&=
	 \sum_{l=0}^\infty t_3
\\\nn&=\frac{-\al^2}{8 \pi  m}
\Bigg[ m \left(3 H_{i m-\frac{1}{2}}+3 \psi ^{(0)}\left(\frac{1}{2}-i m\right)+\gamma +\log (256)\right)
\\\nn& ~~~~~~~~~-i	\ln \Gamma
	\left(\frac{1}{2}-im\right)
 +i \ln \Gamma \left(i m+\frac{1}{2}\right)\Bigg]
}
at $s=-1/2$.

Collecting the contributions from
\Eq{2.3.10a},  \Eq{2.3.15}, \Eq{2.3.19} and \Eq{2.3.20} with \Eq{2.3.14a} and \Eq{2.3.11} we arrive at
\eq{2.3.21}{Z_{2,3}^{reg}&=
\frac{\al^2 }{8 \pi ^{3/2}}
\Bigg[2 \left(4^{s+1}-1\right) s \zeta (2 (s+1)) \sin (\pi  s) \Gamma (-s-1) \Gamma
	\left(s+\frac{3}{2}\right)
\\\nn&~~~~~~~~~
	+\frac{\sqrt{\pi } \left(i \ln \Gamma \left(\frac{1}{2}-i m\right)-i \ln \Gamma \left(i
		m+\frac{1}{2}\right)+m (2 \gamma +\log (16))\right)}{m}\Bigg].
}

\subsubsection{\label{T233}The pole parts and the regular terms from $Z_{23}$}
Here we collect the pole and regular parts from $Z_{23}$, \Eq{2.3.1}.
The complete expression comes from \Eq{2.3.10a},
\Eq{2.3.20}, 
\eq{2.3.22}{
	Z_{2,3} &= Z_{2,3}^{as}+
		Z_{2,3}^{reg}
\\\nn&=
\frac{\al^2 m^{-2 s-1}}{8 \pi ^{3/2}}
\Bigg[
s \sin (\pi  s) \Gamma (-s-1) \Gamma \left(s+\frac{1}{2}\right) \left(\left(4^{s+1}-1\right) (2 s+1)
	m^{2 s+1} \zeta (2 (s+1))+2 s+2\right)
\\&~~~+\sqrt{\pi } m^{2 s} \left(i \ln \Gamma \left(\frac{1}{2}-i m\right)-i \ln\Gamma\left(i m+\frac{1}{2}\right)+m (2 \gamma +\log (16))\right)   \Bigg].
}
Its pole part is,
\eq{2.3.23}{ Z_{2,3}^{pp} &=-\frac{\al^2}{4\pi},
}
and the regular part is
\eq{2.3.24}{ Z_{2,3}^{reg} &=
	\frac{\al^2 \left(i \ln \Gamma \left(\frac{1}{2}-i m\right)-i \ln \Gamma \left(i m+\frac{1}{2}\right)+2 m
		\left(\log \left(\frac{m}{4}\right)+3\right)\right)}{8 \pi  m}.
}
Since these expressions are explicit, it is easy to get their asymptotic expansions. For $m=0$ we get from \Eq{2.3.22}
\eq{2.3.25}{ Z_{2,3}^{m=0} &=
	\frac{\al^2 s\, m^{-2 s-1} \sin (\pi  s) \Gamma (-s-1) \Gamma \left(s+\frac{1}{2}\right) \left(\left(4^{s+1}-1\right) (2 s+1) m^{2
			s+1} \zeta (2 (s+1))+2 s+2\right)}{8 \pi ^{3/2}},
}
and for $m\to\infty$, from \Eq{2.3.24}
\eq{2.3.26}{ Z_{2,3}^{inf} &=
	-\frac{\al^2 (-4 \log (m)-4+\log (16))}{8 \pi }
}
follows.

We recall that in addition we have the 'bad' term, $Z_{2,3}^{bad}$, \Eq{2.3.9}.

\subsection{\label{T24}The part $Z_{3,5}$}

Inserting $h^K_{3,5}$ from \Eq{A1.7} into \Eq{2.8} we get as starting point
\eq{2.4.1}{  Z_{3,5} &= 2\al^3\frac{\sin(\pi s)}{\pi}
	\sum_{l=0}^\infty \nu^{-4}
	\int_{m/\nu}^\infty dz \ ((\nu z)^2-m^2)^{-s}
	\pa_z \frac{-1}{48z^3}.
}
In this case, the $z$-integration can be done,
\eq{2.4.2}{ Z_{3,5}^{'bad'} &= 	\sum_{l=0}^\infty
	\frac{\al^3 m^{-2 s-3} \sin (\pi  s) \Gamma (1-s) \Gamma \left(s+\frac{3}{2}\right)}{12 \pi ^{3/2} \nu
	}.
}
As can be seen, the sum over $\nu$ does not converge and we keep this expression unevaluated. This is the fourth 'bad' term and there are no more contributions from $Z_{3,5}$.

\subsection{\label{T25}The part $Z_{w}$}

Inserting $w(\nu,z)$, \Eq{2.5}, into $Z_w$, \Eq{2.7}, and carrying out the derivative, we get
\eq{2.5.1}{ Z_w &=2\frac{\sin(\pi s)}{\pi}
	\sum_{l=0}^\infty \frac{\al}{2}
	\int_{m/\nu}^\infty dz\, ((\nu z)^2-m^2)^{-s}\frac{1}{z^2}\psi\left(\nu+\frac12+\frac{\al}{2\nu z}\right),
}
where $\psi(x)$ is the logarithmic derivative of the Gamma function. After a substitution $z\to z/\nu$, this expression turns into
\eq{2.5.2}{  Z_w &=2\frac{\sin(\pi s)}{\pi}
	\sum_{l=0}^\infty \frac{\al\nu}{2}
	\int_m^\infty dz\, (z^2-m^2)^{-s}
	\frac{1}{z^2}\psi\left(\nu+\frac12+\frac{\al}{2z}\right).
}
We proceed with a special integral representation of the $\psi$-function,
\eq{2.5.3}{ \psi(x) &= \int_0^\infty dt\,
	\left(x(-\ga-\ln(t))+\frac{1}{t}-\frac{1}{1-e^{-t}}\right)e^{-t x},
}
where $\ga$ is Euler's constant. This representation is a modification of eq.(5.9.13) in \cite{NIST:DLMF}.
With it, $Z_w$ turns into
\eq{2.5.4}{Z_w &= \al\frac{\sin(\pi s)}{\pi}  
	\sum_{l=0}^\infty  \int_0^\infty dt\,\nu e^{-\nu t}\int_{m}^\infty dz \
	g_\nu(z,t)
}
with
\eq{2.5.5}{ g_\nu(z,t) &= \frac{(z^2-m^2)^{-s}}{z^2}
	\left(\left(\nu+\frac12+\frac{\al}{2z}\right)(-\ga-\ln(t))+\frac{1}{t}
	-\frac{1}{1-e^{-t}}   \right)
\exp\left[-t \left(\frac12+\frac{\al}{2z}\right)\right].
}
Now we separate the small $t$ contribution, defining
\eq{2.5.6}{ g_\nu(z,t) &=g_\nu^{as}(z,t)+O(t^3),
}
and split
\eq{2.5.7}{ Z_w &= Z_w^{as}+Z_w^{rerg}
}
with
\eq{2.5.8}{Z_w^{as} &= \al\frac{\sin(\pi s)}{\pi}  
	\sum_{l=0}^\infty  \int_0^\infty dt\,\nu e^{-\nu t}\int_{m}^\infty dz \
	g^{as}_\nu(z,t),
\\\nn 
	Z_w^{reg} &= \al\frac{\sin(\pi s)}{\pi}  
	\sum_{l=0}^\infty  \int_0^\infty dt\,\nu e^{-\nu t}\int_{m}^\infty dz \
	\left(g_\nu(z,t)-g^{as}_\nu(z,t)\right).
}

\subsubsection{\label{T251}The part $Z_w^{as}$}
Here we first integrate over $t$, defining a new function
\eq{2.5.9}{ m_\nu(z) &= \int_0^\infty dt\,e^{-\nu t} 	g^{as}_\nu(z,t),
\\\nn &=
	\frac{\left(z^2-m^2\right)^{-s}}{48 \nu ^3 z^5}
	 \Bigg[6 \left(\left(8 \nu ^3+1\right) z^3+3 z^2+3 z+1\right) 
	 \log (\nu )
	 +\left(24 \nu ^2-35\right) z^2
\\\nn&~~~~~~~~~~~~~~~~~~~~~	 
	+(2 \nu 	-11) z^3	-3 (2 \nu +11) z-9\Bigg].
}
Next we carry out the $z$-integration, which is also explicit, defining
\eq{2.5.10a}{ m_\nu &= \int_m^\infty dz  \ m_\nu(z)
\\\nn	&= \frac{m^{-2 (s+2)}}{192 \sqrt{\pi } \nu ^3}
 \\\nn &~~~~ \times
 	\Bigg[(4 \Gamma (1-s) \Gamma \left(s+\frac{1}{2}\right) 
 \left(m^3 \left(6 \left(8 \nu ^3+1\right) \log (\nu )+2 \nu -11\right)-m
	(2 s+1) (2 \nu -6 \log (\nu )+11)\right)
\\\nn&~~~~~~	+\pi ^{3/2} s \csc (\pi  s) \left(m^2 \left(48 \nu ^2-70\right)+6 \left(6 m^2+s+1\right) \log (\nu )-9
	(s+1)\right)\Bigg].
}

We consider the first three terms in its expansion for large $\nu$. We expand 
\eq{2.5.10}{m_\nu= m_\nu^{as}+O(\nu^{-3}).
}
and get
\eq{2.5.11}{  m_\nu^{as} &=
	\frac{\pi  s m^{-2 s-2} \csc (\pi  s)}{4 \nu }
	+\frac{m^{-2 s-1} \log (\nu ) \Gamma (1-s) \Gamma \left(s+\frac{1}{2}\right)}{\sqrt{\pi
	}}
\\\nn &~~~~	+\frac{\left(m^2-2 s-1\right) m^{-2 s-3} \Gamma (1-s) \Gamma \left(s+\frac{1}{2}\right)}{24 \sqrt{\pi } \nu ^2}.
}
Inserted into $Z_w^{as}$,  \Eq{2.5.8}, these terms  give raise to the last 'bad' terms,
\eq{2.5.12}{Z_w^{bad} &=  \al\frac{\sin(\pi s)}{\pi}  
	\sum_{l=0}^\infty  \nu\, m_\nu^{as},
\\\nn &= \frac{\nu  m^{-2 s-1} \log (\nu ) \sin (\pi  s) \Gamma (1-s) \Gamma \left(s+\frac{1}{2}\right)}{\pi ^{3/2}}
	+\frac{1}{4} s m^{-2 s-2}
\\\nn&~~~~~	-\frac{\left(-m^2+2 s+1\right) m^{-2 s-3} \sin (\pi  s) \Gamma (1-s) \Gamma
	\left(s+\frac{1}{2}\right)}{24 \pi ^{3/2} \nu },
}
since the sum over $\nu$ does not converge.

Now we consider the remaining part of $Z_w^{as}$, i.e., after subtracting $Z_w^{bad}$ and keep the notation $Z_w^{as}$ for this  part,
\eq{2.5.13}{ Z_w^{as} &=        \al\frac{\sin(\pi s)}{\pi}  
		\sum_{l=0}^\infty  \nu\,\left( m_\nu-m_\nu^{as}\right),
\\\nn &=\frac{1}{384} m^{-2 (s+2)}
	 \Bigg[-\pi ^2 s \left(m^2 (-432 \log (A)+70+84 \log (2)+36 \log (\pi ))
\right.\\\nn&\left.	~~~ +(s+1) (-72 \log (A)+9+14 \log (2)
	 +6 \log (\pi ))+6 \gamma  \left(6 m^2+s+1\right)\right)
\\\nn&~~~	 -4 \sqrt{\pi } m (-72 \log
(A)+11+6 \gamma +14 \log (2)
\\\nn&~~~+6 \log (\pi )) \left(m^2
+2 s+1\right) \sin (\pi  s) \Gamma (1-s) \Gamma \left(s+\frac{1}{2}\right)\Bigg].
}
This sum could be done explicitly. It has a pole and a regular part,
\eq{2.5.14}{  Z_w^{as,pp} &=
	\frac{1}{192} \pi  (-72 \log (A)+11+6 \gamma 
	+2 \log (128)+6 \log (\pi )),
\\\nn	 Z_w^{as,reg} &=
	\frac{\pi }{1536 m^3} 
\Big[-16 m^3 (-72 \log (A)+11+6 \gamma +14 \log (2)+6 \log (\pi )) \log (m)
\\\nn&~~~+8 m^3 (\log (4)-2) (-72 \log (A)+11+6 \gamma +14 \log (2)
\\\nn&~~~+6 \log (\pi ))+4 \pi  m^2 (-216 \log (A)+35+18 \gamma +42 \log
		(2)+18 \log (\pi ))
\\\nn&~~~+16 m (-72 \log (A)+11+6 \gamma +14 \log (2)+6 \log (\pi ))
\\\nn&~~~		+\pi  (-72 \log (A)+9+6 \gamma 
	+14 \log (2)+6 \log (\pi ))\Big].
}

\subsubsection{\label{T252}The part $Z_w^{reg}$}
In $Z_w^{reg}$, \Eq{2.5.8}, we change the order and perform the sum over $l$ first. It is explicit and results in
\eq{2.5.15}{p(t,z) &= \sum_{l=0}^\infty \nu e^{-\nu t}
	\left(g_\nu(z,t)-g_\nu^{as}(z,t)\right),
\\\nn&= \frac{ \left(z^2-m^2\right)^{-s}}{-4z^3}\,
	\frac{\tilde p}{t(-1+e^t)^3}
}
where $\tilde p$ is a longer expression and shown in Appendix \ref{E}, eq. \Eq{E.1}.
It is essential that $p(z,t)$ has at $t=0$ a fourth order zero. This allows the integration over $t$, 
\eq{2.5.16}{q(z)&= \int_0^\infty dt\, \frac{\tilde p}{t(-1+e^t)^3},
}
which results also in a longer expression, shown in  Appendix \ref{E}, eq. \Eq{E.2}.

It remains to take the $z$-integration in \Eq{2.5.8},
\eq{2.5.17}{	Z_w^{reg} &= \al\frac{\sin(\pi s)}{\pi}  
\int_{m}^\infty dz \
	 \frac{ \left(z^2-m^2\right)^{-s}}	{-4z^3} \,  q(z),
}
and to take the analytic continuation in $s$. For this we separate the asymptotic part at $z\to\infty$,
\eq{2.5.18}{ q(z) =z\, q^{inf} +O(1)
}
with
\eq{2.5.18a}{
	q^{inf}&=
	\frac{1}{24} 
	\Big[-24 \zeta_H^{(1,0)}(-2,1)
	+24 \zeta_H^{(1,0)}(-2,3)+72 \zeta_H^{(1,0)}(-1,2)-72
	\zeta_H^{(1,0)}(-1,3)
	\\\nn&~~~+24 \zeta_H^{(1,0)}(0,1)
	-72 \zeta_H^{(1,0)}(0,2)
	+48 \zeta_H^{(1,0)}(0,3)
	-48\log (A)-36 \zeta '(2)
	\\\nn&~~~+4+4 \gamma -11 \pi ^2+12 \log (2)-8 \pi ^2 \log (2)\Big]
}
and split
\eq{2.5.19}{ 	Z_w^{reg} &= 	Z_w^{reg,as} + 	Z_w^{reg,reg}
}
with
\eq{2.5.20}{ 	Z_w^{reg,as}&=  \al\frac{\sin(\pi s)}{\pi}  
	\int_{m}^\infty dz \
	\frac{ \left(z^2-m^2\right)^{-s}}	{-4z^2} \,  q^{inf},
\\\nn	Z_w^{reg,reg}&= \al\frac{\sin(\pi s)}{\pi}  
 \int_{m}^\infty dz \
	\frac{ \left(z^2-m^2\right)^{-s}}	{-4z^3} \, (q(z)-z q^{inf}).
}
Again, the integration in the first expression is explicit and delivers
\eq{2.5.21}{Z_w^{reg,as}&=
	\frac{1}{96\pi^{3/2}}\,\alpha\, m^{-1-2s}\,
	\Gamma(1-s)\,\Gamma\!\left(\tfrac{1}{2}+s\right) \\\nn
	&\quad\times\Bigl(-4+\gamma\left(-4+6\pi^2\right)-12\ln 2+48\ln A \\\nn
	&\qquad\qquad +\pi^2\left(11+14\ln 2-72\ln A+6\ln\pi\right)\Bigr)
	\sin(\pi s).
}
It has a pole part,
\eq{2.5.22}{&Z_w^{reg,as,pp}\\\nn&=
-\frac{\al \left(48 \log (A)+\pi ^2 (-72 \log (A)+11+14 \log (2)+6 \log (\pi ))-4+\gamma  \left(6 \pi ^2-4\right)-12 \log (2)\right)}{192 \pi
},
}
and a regular part,
\eq{2.5.23}{\nn	Z_w^{reg,as,reg}&=
	\frac{\al}{192 \pi }
	 \left(48 \log (A)+\pi ^2 (-72 \log (A)+11+14 \log (2)+6 \log (\pi ))-4+\gamma  \left(6 \pi ^2-4\right)
\right.\\ &~~~\left.-12 \log (2)\right) (2 \log
		(m)+2-\log (4)).
}
In contrast to $Z_w^{reg,as,pp}$, the integration in $Z_w^{reg,reg}$ cannot be done explicitly, but it converges for $s=-1/2$,
\eq{2.5.24}{Z_w^{reg,reg}&=
	-\frac{\al}{\pi} \int_m^\infty dz\,\frac{\sqrt{z^2-m^2}}{-4z^3}\,
	(q(z)-z q^{inf})
}
where $q^{inf}$ was defined in \Eq{2.5.18a}.

\subsubsection{\label{T253}The pole parts and the regular terms from $Z_{w}$}
Here we collect the pole and regular parts from $Z_w$. The pole parts result from $Z_w^{as,pp}$, \Eq{2.5.14}, and $Z_w^{reg,as,pp}$, \Eq{2.5.22},
\eq{2.5.25}{Z_w^{pp}&=Z_w^{as,pp}+Z_w^{reg,as,pp},
\\\nn &= \frac{\al (-24 \log (A)+2+2 \gamma +\log (64))}{96 \pi }.
}
The regular parts are $Z_w^{as,reg}$, \Eq{2.5.14},  $Z_w^{reg,as,reg}$, \Eq{2.5.23}, and $Z_w^{reg,reg}$, \Eq{2.5.24}. The first two are explicit and collect into
\eq{2.5.26}{&Z_w^{as,reg}+Z_w^{reg,as,reg}
\\\nn&=
	\frac{\alpha\,\pi}{1536\,m^3}
	\Bigg[\Bigl(11+6\gamma+14\ln 2-72\ln A+6\ln\pi\Bigr)
	\Bigl[\alpha^3\pi+16\alpha^2 m+8m^3(\ln 4-2)-16m^3\ln m\Bigr] \\\nn
	&~~~~~~~~~~~~~~~~~~+ 4\alpha m^2\pi
	\Bigl(35+18\gamma+42\ln 2-216\ln A+18\ln\pi\Bigr) \Bigg]\\\nn
	&\quad ~~~~ + \frac{1}{192\,\pi}\,\alpha\bigl(2-\ln 4+2\ln m\bigr)
\\\nn&	~~~~~~~~~~~~~~~~~~\cdot\Bigl(-4+\gamma(-4+6\pi^2)-12\ln 2+48\ln A+\pi^2(11+14\ln 2-72\ln A+6\ln\pi)\Bigr),
}
whereas the third,  $Z_w^{reg,reg}$, \Eq{2.5.24}, is left for numeric evaluation.

For $m\to\infty$ we have a non-decreasing contribution only from $Z_w^{as,reg}+Z_w^{reg,as,reg}$ and it is
\eq{2.5.27}{Z_w^{inf}=-\frac{\al (-12 \log (A)+1+\gamma +\log (8)) \left(\log \left(\frac{m}{2}\right)+1\right)}{24 \pi }.
}

For $m\to0$ we observe a special behavior. While all   contributions to the zeta function from the preceding subsections are finite in this limit, here we observe a singular behavior from $Z_w^{as,reg}$, \Eq{2.5.14}, and from
$Z_w^{reg,reg}$, \Eq{2.5.20}. While $Z_w^{as,reg}$ is explicit, for $m\to 0$, for $Z_w^{reg,reg}$ we need to consider the integration over $z$ which becomes singular at $z\to0$. This can be seen from the expansion of $q(z)$, \Eq{E.2} for small $z$,
\eq{2.5.28}{
 q(z)&= \frac{-36 \zeta '(2)-9 \pi ^2-8 \pi ^2 \log (2)}{24 z^2}\al^3
\\\nn&~~~+\frac{-36 \zeta '(2)+4 \log (z)-11 \pi ^2-4 \gamma +2
	-8 \pi ^2 \log (2)-4 \log	(2)}{8 z}\al^2
\\\nn&~~~+\frac{1}{24} \left(-108 \zeta '(2)-35 \pi ^2
	-24 \pi ^2 \log (2)\right)\al
+O\left(z^1\right),
}
which can be obtained by an expansion of the Hurwitz zeta functions and the Stieltjes numbers in \Eq{E.2} for a large parameter.
Inserted into $Z_w^{reg,reg}$, \Eq{2.5.24}, and added to the expansion of $Z_w^{as,reg}$, the odd powers of $m$ compensate, leaving
\eq{2.5.29}{Z_w &=
	-\frac{\al^3 (6 \log (\al)+6 \gamma -11+\log (4096))}{144 \pi  m^2} +O(1)
}
as the leading contribution for $m\to 0$.

\subsection{\label{T26}Collecting the 'bad' terms}
Here we collect the 'bad' terms to demonstrate how these cancel. The first one comes from $Z^{as}_{1,1}$, \Eq{2.1.5},
\eq{2.6.1}{  Z_{1,1}^{as} &=
	\sum_{l=0}^\infty
	\frac{-\al \nu  m^{-2 s-1} \log (\nu ) \sin (\pi  s) \Gamma (1-s) \Gamma \left(s+\frac{1}{2}\right)}{\pi ^{3/2}}.
}
The second comes from $Z_{1,3}$, \Eq{2.2.3},
\eq{2.6.2}{ Z_{1,3} &= \sum_{l=0}^\infty
	\frac{-\al m^{-2 s-1} \sin (\pi  s) \Gamma (1-s) \Gamma \left(s+\frac{1}{2}\right)}{24 \pi ^{3/2} \nu },
}
the third, $Z_{2,3}^{as}$, first term in \Eq{2.3.8}, is
\eq{2.6.3}{  Z_{2,3}^{as} &=	\sum_{l=0}^\infty
	\frac{-\al^2}{8} \pi  s m^{-2
		s-2},
}
the fourth, $Z_{2,5}$, \Eq{2.4.2}, is
\eq{2.6.4}{ Z_{3,5} &= 
	\sum_{l=0}^\infty
\frac{\al^3 m^{-2 s-3} \sin (\pi  s) \Gamma (1-s) \Gamma \left(s+\frac{3}{2}\right)}{12 \pi ^{3/2} \nu },
}
and finally the last is $Z_w^{as,as}$, \Eq{2.5.12},
\eq{2.6.5}{ Z_w^{as,as} &=  \sum_{l=0}^\infty
\Bigg[ 
\frac{\al \nu  m^{-2 s-1} \log (\nu ) \sin (\pi  s)
	\Gamma (1-s) \Gamma \left(s+\frac{1}{2}\right)}{\pi ^{3/2}}
\\\nn&~~~~~~~~~-\frac{\al m^{-2 s-3} \sin (\pi  s) \Gamma (1-s) \Gamma
	 \left(s+\frac{1}{2}\right) \left(2 \al^2
	s+\al^2-m^2\right)}{24 \pi ^{3/2} \nu }
+\frac{1}{4} \al^2 s m^{-2 s-2}
\Bigg].
}
As can be seen, these terms compensate when added together. We mention that just this compensation was one of the reasons not to expand the full asymptotic expansion of the Kummer function for all arguments large, for large $\nu$ (in our notations), but to keep $w(\nu,z)$ in eq. \Eq{2.3} unexpanded and to include $k=5$ into $\ln\Phi^{as}(\nu,\nu z)$.

\subsection{\label{T27}Collecting pole and regular parts of the Kummer contribution to the asymptotic part of the  zeta function}

In this subsection we collect the Kummer contributions to the asymptotic part of the zeta function \Eq{i13}, which were calculated in the preceding subsections. The analytic continuation to $s=-1/2$ shows a double pole, a single pole and a regular contribution.

The double pole results from $Z_{1,1}^{ppp}$, \Eq{2.1.21}, and from $Z_{1,3}^{ppp}$, \Eq{2.2.14}, which compensate each other and   therefore it is absent in the final expression.

The simple pole results from
$Z_{1,1}^{pp}$, \Eq{2.1.21},
$Z_{1,3}^{pp}$, \Eq{2.2.14},
$Z_{2,3}^{pp}$, \Eq{2.3.23},
$Z_{w}^{pp}$, \Eq{2.5.25},
\eq{2.7.2}{ Z_K^{pp}&= -\frac{\al^2}{4\pi}+\frac{\al}{48 \pi ^3}
	 \Big[-3 \ga_1^{(0,1)}\left(1,\frac{1}{2}\right)+3 \ga_1^{(0,1)}(1,1)
\\\nn &~~~~
	 +\pi ^2 \left(-12
		\log (A)-6 m^2+\gamma +3+\log (\pi )+\log (8)\right)\Big]
}
where $Z_K^{pp}$ denotes the contribution from the Kummer-terms and $\ga_1$ was defined in \Eq{2.1.13c}.

Also we collect the contributions not decreasing for $m\to\infty$.
These result from
$Z_{1,1}^{inf}$, \Eq{2.1.23},
$Z_{1,2}^{inf}$, \Eq{2.2b.15},
$Z_{1,3}^{inf}$, \Eq{2.2.17},
$Z_{2,3}^{inf}$, \Eq{2.3.26},
$Z_{w}^{inf}$, \Eq{2.5.27},     and collect into
\eq{2.7.3}{ Z_K^{inf}&=
	\frac{\al}{3840 \pi ^3}
\\\nn & ~~~\times	 \Bigg[-160 \left(\log \left(\frac{m}{2}\right) \left(3 \ga_1^{(0,1)}(1,1)+\pi ^2 (-12 \log (A)-12
		\al+\gamma +\log (\pi )+\log (8))\right)
\right. \\\nn & ~~~~~\left.
		+\ga_1^{(0,1)}\left(1,\frac{1}{2}\right) (\log (8)-3 \log (m))+\pi
		^2 \left(\left(3-6 m^2\right) \log (m)+\log (\pi )\right)\right)
\\\nn&~~~
	+480
		\left(\ga_1^{(0,1)}\left(1,\frac{1}{2}\right)-\ga_1^{(0,1)}(1,1)\right)
\\\nn&~~~
		-160 \pi ^2 (-12 \log
		(A)-12 \al+m (m (\log (64)-3)-6 \pi )+\gamma +3)
		\Bigg].
}
Also we collect the contributions at $m=0$, but general $s$. These result from
$Z_{1,1}^{zero}$, \Eq{2.1.22},
$Z_{1,2}^{zero}$, \Eq{2.2b.16},
$Z_{1,3}^{zero}$, \Eq{2.2.18},
$Z_{2,3}^{zero}$, \Eq{2.3.25},

 and collect into
\eq{2.7.4}{Z^{zero} &=
	\frac{\al }{576 \pi ^{3/2}}
	\Bigg[72 s \sin (\pi  s) \Gamma (-s-1) \Gamma \left(s+\frac{1}{2}\right)
\\\nn & ~~~
 \left(2 \al (s+1) m^{-2
			s-1}+\al \left(4^{s+1}-1\right) (2 s+1) \zeta (2 (s+1))-\frac{8 \left(4^s-1\right) (s+1) \zeta (2 s)}{2
			s+1}\right)
\\\nn & ~~~
	+\sqrt{\pi } \left(24 \left(\gamma _1+3+\log (512)\right)+\pi ^2-84 \log ^2(2)-72 \gamma  (\log
		(2)-1)\right)\Bigg].
}

\section{\label{T3}The Kummer part of the heat kernel coefficients and their contribution to the  divergent part of the Zeta function}

The heat kernel for the scalar field in a Coulomb background is defined in \Eq{i10} in terms of the heat kernel coefficients $a_n$. In our case with a Coulomb background and a boundary, only the Bessel part $a^B_n$ in
\eq{3.1}{a_n&= a_n^B +a_n^K,
}
is known. These are just the contributions without $\al$,
\eq{3.2}{ a^B_n &={a_n}_{|{_{\al=0}}},
}
and these are well known, see for instance eq.(15) in \cite{bord26-8-79}. We reprinted them in Appendix \ref{Ahk}, eq. \Eq{hk}.
The Kummer contribution is proportional to $\al$ and, as it turned out,  to  $\al^2$, and it is not known.

The relation between the heat kernel coefficients and the divergent part of the zeta function is shown in \Eq{i11}. It consists of the pole parts and the non-decreasing at $m\to\infty$ contributions,
\eq{3.3}{\zeta^{div}_K(s) &=
	\frac{Z_K^{pp}}{s+1/2}+Z_K^{inf}.
}
These were calculated in the preceding section, \Eq{2.7.2} and \Eq{2.7.3}.

From comparing representations \Eq{i11} and \Eq{3.3} of the divergent part of the zeta function the coefficients can be read off,
\eq{3.5}{ a_2^K &=
	2 \pi  \al^2
	-\frac{\al}{3 \pi	}
	\Bigg[-3 \ga_1^{(0,1)}\left(1,\frac{1}{2}\right)+3
	\ga_1^{(0,1)}(1,1)-12 \pi ^2 \log (A)
	+\gamma  \pi ^2+3 \pi ^2
	\\\nn &~~~~~~~~	~~~~~	~~~~~~~~
	+\pi ^2 \log (\pi )+\pi ^2 \log (8)\Bigg],
\\\nn  a_{3/2}^K&= 2\pi^{3/2} \al,
\\\nn
	a_1^K &= -2\pi \al,
\\\nn a_{1/2}^K &=0.
}
With these coefficients, equations \Eq{i11} and \Eq{3.3} become identical.

 A second way to get the coefficients follows from the relation between the zeta function and the heat kernel at $m=0$, see for example, eq.(2.9) in \cite{bord26-8-79}, and delivers with
\eq{3.4}{ a_n^K &= (4\pi)^{3/2}{\rm Res}_ {s=3/2-n}
	\Gamma(s)\zeta(s)_{|_{m=0}}.
}
a direct way to calculate the coefficients. The corresponding zeta function was calculated in \Eq{2.7.4} and confirms just the coefficients \Eq{3.5}. It serves as an internal consistence check. Another, partial, check is the comparison with the usual bulk contributions to the coefficients, $a_1=-\int_{r<R} V(x)dx$ and $a_2=\frac12\int_{r<R} V(x)^2dx$, which coincide with the corresponding fragments in \eqref{3.5}.

\section{\label{T4}The analytic part of the zeta function}

According to \Eq{i17}, after subtracting the divergent part from the asymptotic part, we get the analytic part. It was calculated in section \ref{T2} and its individual parts were denoted as regular parts.

These regular contributions come from
$Z_{1,1}^{reg}, \Eq{2.1.21}$,
$Z_{1,2}^{reg}, \Eq{2.2b.9}$,
$Z_{1,3}^{reg}, \Eq{2.2.16b}$,
$Z_{2,3}^{reg}, \Eq{2.3.24}$,
$Z_{w}^{reg},  \Eq{2.5.28}$,
and consist of an explicit part, and a numerical part,
\eq{3.9a}{\zeta^{reg}=\zeta^{reg,an}+\zeta^{reg,num}.
}

\begin{figure}[h]
	\includegraphics[width=16 cm]{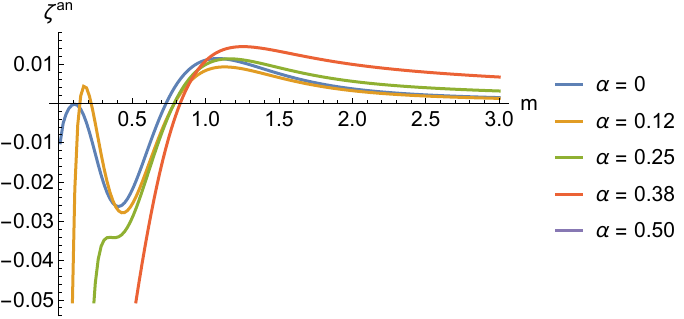}
	\caption{The functions $\zeta_K^{an}$, \Eq{3.11}, as a function on $m$ for several values of $\al$.
	}		\label{fig:1}
\end{figure}

The explicit part, $\zeta^{reg,expl}$, is
\eq{3.8}{\zeta^{reg,an} &=
	-\frac{\al}{576 \pi ^3 m}
	\Bigg[\pi ^2 m \left(12 \log (m) \left(\log \left(16 \pi ^2 m\right)-12 \left(\al+m^2\right)\right)+144
	\al (\log (4)-3)
	\right. \\\nn & ~~~\left.
	+72 m^2 (\log (4)-1)+24 \gamma  \log (4 m)+\pi ^2+24 \left(-3+\log (2) \log \left(\frac{2}{\pi
	}\right)+\log \left(\frac{\pi }{4}\right)\right)\right)
	\\\nn&~~~
	+12 \left(6 m
	\left(\ga_1^{(0,1)}(1,1)-\ga_1^{(0,1)}\left(1,\frac{1}{2}\right)\right) \left(\log
	\left(\frac{m}{2}\right)+1\right)
	-6 i \pi ^2 \al \ln \Gamma \left(\frac{1}{2}-i m \right)
	\right. \\\nn & ~~~\left.
	+6 i \pi ^2 \al
	\ln \Gamma \left(i m+\frac{1}{2}\right)+\pi ^2 m \left(2 \psi ^{(0)}\left(\frac{1}{2}
	-i m\right)+2 \psi ^{(0)}\left(i	m+\frac{1}{2}\right)
	\right.\right. \\\nn & ~~~\left.	\left.
	-5 i m \left(\psi ^{(1)}\left(\frac{1}{2}-i m\right)-\psi ^{(1)}\left(i m+\frac{1}{2}\right)\right)-2
	\gamma _1\right)\right)\Bigg]
	\\\nn&~~~
				+\frac{\pi  \al}{1536 m^3}
		\Bigg[\pi  \al^3 (-72 \log (A)+9+6 \gamma +14 \log (2)
	+6 \log (\pi ))
	\\\nn&~~~~~~~~~~~~~~~~~~~~
	+16 \al^2 m (-72 \log
	(A)+11+6 \gamma +14 \log (2)+6 \log (\pi ))
	\\\nn&~~~~~~~~~~~~~~~~~~~~
	+4 \pi  \al m^2 (-216 \log (A)+35+18 \gamma +42 \log (2)+18 \log (\pi ))
	\\\nn&~~~~~~~~~~~~~~~~~~~~
	-16
	m^3 (-72 \log (A)+11+6 \gamma +14 \log (2)+6 \log (\pi )) \log (m)
	\\\nn&~~~~~~~~~~~~~~~~~~~~
	+8 m^3 (\log (4)-2) (-72 \log (A)+11+6 \gamma +14 \log (2)+6
	\log (\pi ))\Bigg],
}
where $\ga_1$ is defined in \Eq{2.1.13c}.
The numerical part is
\eq{3.9}{   \zeta^{reg,num} &= \al \sum_{l=0}^\infty
	\left[\frac{1}{4} \left(1-\frac{1}{\sqrt{\frac{m^2}{\nu ^2}+1}}\right)+\frac{\log \left(\frac{m^2}{\nu ^2}+1\right)}{48 \pi  \nu }   \right]
	\\\nn&
	+\frac{2\al}{\pi}
	\int_m^\infty dz\,\frac{1}{\sqrt{z^2-m^2}}\,
	\\\nn& ~~~ \times	\left\{
	\int_0^z \frac{d\nu\ \nu}{1+\exp(2\pi \nu)}
	\ln\left[\frac12\left(1+\sqrt{1-\left(\frac{\nu}{z}
		\right)}\right)\right]
	-	\int_z^\infty \frac{d\nu\ \nu}{1+\exp(2\pi \nu)}
	\ln\frac{2z}{\nu}\right\}.
}
Returning to the notation of equation \Eq{i17}, we have with
\eq{3.11}{\zeta^{an} &= \zeta^{reg}
}
with \Eq{3.9a} an expression which can be calculated numerically and plotted.
This had been done in Figure \ref{fig:1}, which shows $\zeta^{an}$ as a function of $m$ for several values of $\al$ . For $\al=0$ we reproduce the dashed curve in Figure 2 in \cite{bord26-8-79}. For increasing $\al$, the Kummer contributions grow fast in magnitude, and for instance, become singular at $m\to0$.

\section{\label{T5}The numerical part of the   Zeta function}

In this section we consider the numerical part of the zeta function which was defined in \Eq{i14}. Written in detail, and similar to \Eq{2.2}, it reads
\eq{4.1}{ \zeta^{num}(s) &=
	\lim\limits_{s\to-1/2}
	2 \frac{\sin(\pi s)}{\pi}\sum_\nu \nu\int_{m/\nu}^\infty\, dz\,
	\left[(\nu z)^2-m^2\right]^{-s}\pa_z
	\left(\ln\Phi_l(i\nu z)-\ln\Phi^{as}_l(i\nu z)\right)
}
with $\Phi^{as}_l(i\nu z)$ defined in \Eq{2.3}.
We mention that the difference in the parentheses vanishes for large $\nu$ as $\nu^{-4}$ and for large $z$ as $z^{-4}$.
Since we want for consistency also to consider the limiting case $\al\to0$ (Bessel case) and the calculations for that case were done in another paper, we divide the asymptotic part of the logarithm of the zeta function into two pieces,
\eq{4.2}{ \ln \Phi^{as}(\nu,i\nu z) &= F_B +F_K,
}
where   Bessel  and Kummer parts are given by
\eq{4.3}{F_B &= \sum_{k=-1}^3 \nu^{-k}h_{0,k}(z),
	&F_K &= \sum_{p=1}^2\sum_{k=0}^3  \al^p \nu^{-k}h_{p,k}(z)
		+\al^3\nu^{-5}h_{3,5}(z)+w(\nu,i\nu z),
}
where we also accounted for \Eq{2.4}. We remind that the coefficients for $k=-1,0$ are pure Bessel, as mentioned in Appendix \ref{A}.

To establish the connection to the Bessel case we consider the limit $\al\to0$,
\eq{4.4}{\lim\limits_{\al\to0}
	\ln\Phi_l(i\nu z)&=
\nu\ln2+\ln\Gamma(\nu+1)+\ln\Phi_l^B(i\nu z)
}
where, in terms of a modified Bessel function,
\eq{4.5}{\ln\Phi_l^B(i\nu z) &=\ln\left((\nu z)^{-\nu}I_\nu(\nu z)\right)
}
holds.

The coefficients for the Bessel functions are well known and we take them from the recent \cite{bord26-8-79}, eq. (36). The resulting expression we denote by
\eq{4.6}{\tilde F_B &=\sum_{k=-1}^3\nu^{-k}D_k(t).
}
With the coefficients from Appendix \ref{A} the following relations hold,
\eq{4.7}{F_B &= -\nu\ln(2\pi\nu)+\ln(\pi/2)+\tilde F_B.
}
This way, both expansions, \Eq{4.4} and \Eq{4.7} differ by a constant, which drops out under the differentiation in eq. \Eq{4.1} and we can use the results from \cite{bord26-8-79} for both, the Bessel parts of the logarithm and its expansion, as well as for the asymptotic part of the zeta function, eq.(33) in that paper.

For the numerical part of the zeta function we get in these notation
\eq{4.8}{ \zeta^{num}  &=
	\lim\limits_{s\to-1/2}
	2 \frac{\sin(\pi s)}{\pi}\sum_\nu \nu\int_{m/\nu}^\infty\, dz\,
	\left[(\nu z)^2-m^2\right]^{-s}
\\\nn & ~~~~  \times
	\pa_z
	\left[\ln\Phi(\nu,i\nu z)
	-\left(-\nu\ln(2\nu z)+\frac12\ln(\pi)+\tilde F_B+F_K\right)\right].
}
We mention that the coefficients in $\tilde F_B$ are defined for the Bessel function, i.e., without the factor $(\nu z)^{-\nu}$ in \Eq{4.5}.
The asymptotic part has the property that the difference in the square bracket in \Eq{4.8} decreases fast enough to make the sum/integral converge. For this reason one can carry out the limit and arrives at
\eq{4.9}{ \zeta^{num}  &=
-\frac{2}{\pi}\sum_\nu \nu\int_{m/\nu}^\infty\, dz\,
\sqrt{ (\nu z)^2-m^2}
	\\\nn & ~~~~  \times
	\pa_z
	\left[\ln\Phi(\nu,i\nu z)
	-\left(-\nu\ln(2\nu z)+\frac12\ln(\pi)+\tilde F_B+F_K\right)\right].
}
However, for numerical evaluation the derivative is inconvenient. Therefore we integrate by parts,
\eq{4.10}{ \zeta^{num} &=
	\frac{2}{\pi}\sum_\nu \nu\int_{m/\nu}^\infty\, dz\,
	\frac{\nu^2 z}{\sqrt{ (\nu z)^2-m^2}}
	\\\nn & ~~~~  \times
	\left[\ln\Phi(\nu,i\nu z)
	-\left(-\nu\ln(2\nu z)+\frac12\ln(\pi)+\tilde F_B+F_K\right)\right].
}
There are no boundary terms since the integrand vanishes for $z=m/\nu$ from the square root in \Eq{4.9} and from  the decrease for large $z$ at infinity.

For the numerical work we represent $ \zeta^{num}$ as a sum
\eq{4.11}{	 \zeta^{num}
	=\sum_{l=0}^\infty \nu \ \zeta^{num}_l
}
with
\eq{4.12}{ \zeta^{num}_l &=   \frac{2}{\pi}
\int_{m/\nu}^\infty\, dz\,
	\frac{\nu^2 z}{\sqrt{ (\nu z)^2-m^2}}
	\\\nn & ~~~~  \times
	\left[\ln\Phi(\nu,i\nu z)
	-\left(-\nu\ln(2\nu z)+\frac12\ln(\pi)+\tilde F_B+F_K\right)\right].
}
As mentioned, the integral is fast converging. After integration, the sum is also convergent, $\zeta^{num}_l$ decreases as $\nu^{-3}$.

\begin{figure}[h]
	\includegraphics[width=16 cm]{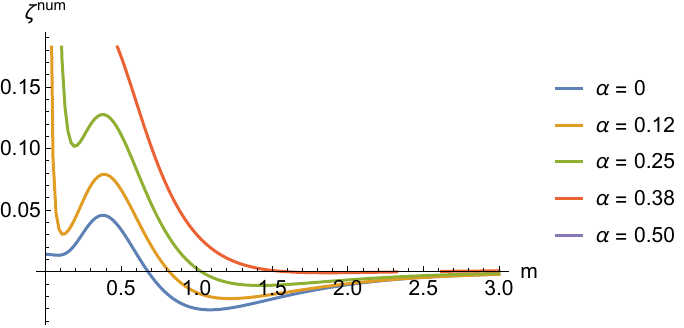}
	\caption{The functions $\zeta^{num}$, \Eq{4.11}, as a function on $m$ for several values of $\al$.
	}		\label{fig:2}
\end{figure}

 We plot $\zeta^{num}$ in Figure \ref{fig:2} for several values of the coupling $\al$.  For  $\al=0$ we reproduce the corresponding curve in Figure 1 in \cite{bord26-8-79}. However, as can be seen, for increasing $\al$ the Kummer contributions soon dominate. For $m\to0$ they diverge.

\section{\label{T6}The renormalized  Zeta function}

\begin{figure}[h]
	\includegraphics[width=16 cm]{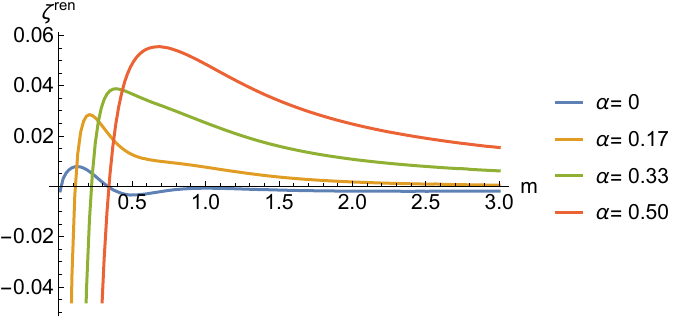}
	\caption{The functions $\zeta_K^{ren}$, \Eq{5.1}, as a function on $m$ for several values of $\al$.
	}		\label{fig:3}
\end{figure}

Inserting \Eq{i17} into \Eq{i16} we get the renormalized zeta function in the form
\eq{5.1}{\zeta^{ren}&=\zeta^{num}+\zeta^{an}
}
and using \Eq{4.12} and \Eq{3.11} we can plot the renormalized zeta function.
The result is shown in Figure \ref{fig:3} as a function of $m$ for several values of $\al$. Again, this zeta function grows fast with the coupling $\al$ of the Coulomb background field.

For small mass it grows as $m^{-2}$, except for $\al=0$, where we observe the finite value at $m=0$ from the Bessel case.The responsible contributions result from $Z_w$, \Eq{2.5.29}, and from $\zeta^{num}$, \Eq{4.1}.
From $Z_w$ we have an explicit expression,
  \eq{5.2}{\zeta^{an}&\simeq
 	\frac{\al^3}{m^2}\Big[
 	\frac{1}{96} \pi  (-72 \log (A)+11+6 \gamma +14 \log (2)+6 \log (\pi ))
 \\\nn&~~~~~~~~~~~-\frac{12 \log (\al)+108 \zeta '(2)
 	+12 \gamma -22+3 \pi ^2 (11+\log (256))+24 \log (2)}{288 \pi }\Big],
 \\\nn \sim &~	\frac{\al^3}{m^2}\left( -0.432981-0.0265258 \ln\al\right),
 }
 
 where the first term dominates (until very low $\al$).
Together with  $\zeta^{num}$  the sign of the vacuum energy is negative for  small $m$ and all $\al$.

For large mass we have
\eq{5.3}{\zeta^{an}=-\frac{\al}{96\pi m}
	  +\frac{\al^2\pi^2}{384m}
	\left(-216 \log (A)+35+18 \gamma +42 \log (2)+18 \log (\pi )\right)
	+O(m^{-2}).
}
This contribution changes sign at $\al\simeq0.0098$ and dominates the asymptotic in $\zeta^{ren}$ for larger $\al$. The contribution from $\zeta^{num}$ decreases faster than $m^{-2}$.

\section{\label{TC}Conclusions}
We calculated the zeta function $\zeta(s)$ for the Laplace operator with background potential
\eq{c.1}{V^{backgr} &=\frac{\al}{r}
}
with a massive, neutral scalar field and Dirichlet boundary conditions at $r=R$ in 3 dimensions. It provides the zetafunctional regularization  of the vacuum energy \Eq{i3}. The renormalization is performed by subtraction of $\zeta^{div}$, \Eq{i11}.  Technically, for the calculations we used methods which generalize those developed in \cite{bord97-56-4896} and rechecked recently in \cite{bord26-8-79}. A key moment is the normalization condition used, and explained in the Introduction, which ensured that the renormalized vacuum energy $E_0^{ren}$ behaves for large mass $m$ as
\eq{c.2}{E_0^{ren} &\sim \frac{1}{m},
}
which we consider as physical normalization condition. The counterterms, which follow from the zeta function in this approach from eq. \Eq{3.5}, have contributions proportional to $\al$ and to $\al^2$.

Returning to the     system with a spherical surface ("bag") with classical energy,
\eq{c.3}{E_{class}&= pV+\sigma S+k+F R+\frac{h}{R},
}
introduced in \cite{bord97-56-4896}, eq. (1), for an empty sphere, we would conclude that the divergent term proportional to $\al$ (it has the dimension of a length) renormalizes $F$ and that proportional to $\al^2$ renormalizes $\sigma$, the surface tension. For comparison we display the heat kernel coefficients for the old case, i.e., for $\al=0$,  in Appendix \ref{Ahk}, eq. \Eq{hk}. It must be underlined, that this system is a pure model. In case of an electric charge in place of $\al$ and the electromagnetic field we would have in addition the infinite electrostatic self energy of the charge. In case of gravity, we would have a black hole in place of a point mass. It is interesting to mention that we observe here a parallel to semiclassical gravity, where also a counterterm quadratic in the coupling, the curvature in that case, appears. This is simply a consequence of the circumstance that all counterterms which are allowed by dimension may indeed appear.

\begin{figure}[t]
	\includegraphics[width=16 cm]{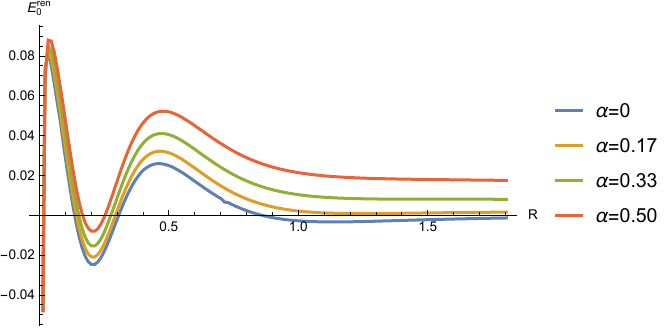}
	\caption{The  vacuum energy $E_0^{ren}$, \Eq{i18a}, as a function on $R$ for several values of $\al$. The mass is $m=1$.
	}		\label{fig:4}
\end{figure}

\begin{figure}[h]
	\includegraphics[width=9 cm]{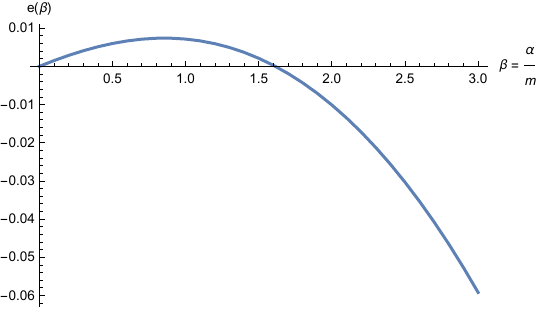}
	\caption{The  function $e(\beta)$ in eq. \Eq{c6} as function of its argument  $\beta=\frac{\al}{m}$. 
	}		\label{fig:5}
\end{figure}

The renormalized vacuum energy $E_0^{ren}$ follows from the renormalized zeta function by eq. \Eq{i18a}. That is, the curves in Figure \ref{fig:3}, rescaled according to \Eq{i18} and divided by R. The result is shown in Figure \ref{fig:4} as function of $R$. The lowest curve ($\al=0$) repeats the curve in Figure 3 in \cite{bord26-8-79}. The singularity at $R\to0$ is $\sim1/R$ as expected from dimensional reasons. Increasing $\al$ lifts the curves. At $R\to\infty$, the behavior changes. While the energy goes to zero for $\al=0$, i.e. for the Bessel case, with non-zero $\al$, i.e., with Coulomb background, it tends to a constant,
\eq{c4}{ E_0^{ren} & \raisebox{-1.ex}{\ensuremath{\stackrel{\to}{\phantom{}_{R\to\infty}}}}
 \, E_0^{ren,inf}(\al/m),
}
where
\eq{c6}{ E_0^{ren,inf}(\al,m) &= \al \, e(\al/m)
}
and $e(\al/m)$ is a dimensionless function. It is shown in Figure \ref{fig:5} as a function of its argument. We mention that $e(\al/m)$ results completely from $Z_w^{as,reg}$, \Eq{2.5.14}, and from  $Z_w^{reg,as}$, \Eq{2.5.20}. Since we by our renormalization prescription kept only contributions decreasing at least as $m^{-1}$, for that limit we need $\al^2$ and higher. So the only other candidate is $Z_{2,3}$, \Eq{2.3.1},, which however, decreases as $m^{-2}$.
 All other contributions, especially $\zeta^{num}$, \Eq{4.1}, are smaller and also do not contribute to \Eq{c6}.

The existence of this limit, independent of $R$, shows that a finite renormalized vacuum energy can be defined for a massive scalar field in the background of a Coulomb source. For $m\to\infty$ it goes to zero in accordance with our normalization condition and for $m\to0$ it diverges, demonstrating that the given calculation cannot be applied to a massless field. The existence of the finite vacuum energy\Eq{c6}, however, does not necessarily imply the existence of  a well defined vacuum energy defined initially in the infinite volume.
We mention that two limits are involved here. First, $s \to 0$ (removing the regularization), and afterwards $R_- \to \infty$ (removing the Dirichlet sphere). The reverse order, i.e., starting from infinite volume, encounters the problem of the logarithmic Coulomb phase, even for a massive field. This is presumably the reason why a result such as \Eq{c6}  could not be obtained so far.

Technically, the present calculation is interesting  as it went quite far beyond the methods known from the Bessel case. The key moment is the impossibility to use the complete uniform asymptotic expansion for large $\nu$ of the mode generating function, as explained in section \ref{T2}. It was necessary to keep a part not completely expanded. However, using the integral representation \Eq{2.5.3} for the derivative of the logarithm of the Gamma function, it was possible to proceed. Regrettably, there is no suitable integral representation directly for the logarithm of the Kummer function.

The Casimir effect, as another special case of the vacuum energy, is by now well established and allows precision measurements. However, physical applications are so far restricted to forces acting between separate bodies. There is no physical example for the vacuum energy of a single body. The current calculation is a step towards a physical example.

If extended to a single charge in electrodynamics, due to the finite volume one could avoid the problem of the infrared photons, which inevitably accompany an electron in infinite space, and hampering the integration of the energy density due to the Coulomb phase. As is known, such a photon cloud changes the vacuum state away from that of a single electron. The same will happen for a scalar field if it is massless. This is supported by the singular limit $m\to0$ seen  in this paper.

Another problem is that in our calculation the source of the external field is pointlike. However, a inner sphere would not change the calculation principally, but only technically. Basically, one had an additional contribution in terms of Tricomi functions in addition to the  Kummer functions. When changing the interior region accordingly, one could also consider the vacuum energy of a black hole. However, a vacuum energy calculated from the quasi normal modes would probably have  an imaginary part.

\appendix
\section{\label{A}The coefficients of the asymptotic expansion of the Kummer function $M(a,b;z)$}
In this section we derive the functions $h_k(z)$ and $w(\nu,z)$ appearing in eq. \Eq{2.3}.
For this we need the asymptotic expansion of the Kummer function for all arguments  large. It was obtained in \cite{temm22-33-1221}.

Theorem 3.1 states that the uniform asymptotic expansion reads
\eq{A1.1}{ \frac{1}{\Ga(b)}{_1}F_1(a,b,c) &\sim e^{z(1-A)}z^{a-b}\frac{f_0}{\Ga(a)}\sum_{n\ge0}\frac{\tilde f_n}{z^n},
}
where
\eq{A1.2}{
	\mu&=\frac{b-a}{z},
	\be=\frac{b}{z},
	\al_T=\frac{a}{z},
	&\tau &=\frac{2}{\be+1+\sqrt{(\be+1)^2-4\mu}},
\\\nn
	 A&=\mu(\tau-\ln\tau-1)-\al_T\ln(1-\mu\tau),
	 &f_0 &= \frac{1}{\sqrt{\be\mu\tau^2-2\mu\tau+1}},
 	& \tilde f_0 &=1.
}
The first two coefficients $\tilde f_n$   were shown in the paper. Two more coefficients were provided by N.Temme\footnote{Private communication} and are displayed in Appendix B. 
Inserting the variables
\eq{A1.3}{ a=l+1+\frac{\al}{2\nu\xi}, \
		b=2l+2, \
		z=2\nu\xi ,
}
appearing in \Eq{i7a}, and re-expanding for large $\nu$ all terms except for  $w(\nu,z)$ defined in  \Eq{2.3} and \Eq{2.5}, we get
\eq{A1.4}{ \ln \Phi^{as} &=
	\sum_{k=-1}^3  \frac{1}{\nu^k}h_k(z)+\frac{1}{\nu^5}h_5(z)+w(\nu,z).
}
The coefficients $h_k(z)$ are polynomials in the coupling $\al$. Contributions without $\al$ are the same as for the modified Bessel function $I_\nu(z)$. Beyond the Bessel contributions we find for the z-derivatives
\eq{A1.5}{
\pa_z h_1(z) &=  -\frac{\log (\nu )-\frac{z}{\sqrt{z^2+1}}+\sinh ^{-1}(z)}{2 z^2}\al,
\\\nn
\pa_z h_2(z) &=   \frac{z}{2 \left(z^2+1\right)^2}\al,
\\\nn
\pa_z h_3(z) &= \frac{z \left(z^4+26 z^2-50\right)}{48 \left(z^2+1\right)^{7/2}}\al
	+\frac{-3 z^2-2}{8 z^3 \left(z^2+1\right)^{3/2}}\al^2.
}
This way the Kummer contribution starts from next-to-next to leading in the asymptotic expansion for large $\nu$.
 
In \cite{bord95-179-215} (and later papers), the calculation of the asymptotic contributions to the Bessel functions was done in terms of the variable
\eq{A1.6}{  t &= \frac{1}{\sqrt{1+z^2}}.
}
 We mention that the above re-expansion of the coefficients from Appendix \ref{B} is a nontrivial task; it was done in terms of $t$ for the derivatives, and resulted in \Eq{A1.2}, where it was re-displayed in the variable $z$.
Integrating back and  using the variable $t$,
we get for the complete contributions, i.e., including the Bessel case for $\al=0$,
\eq{A1.7}{
h_1(t) &=
	\frac{1}{24} t \left(3-5 t^2\right)
	+ \frac{t \left(\sinh ^{-1}\left(\frac{\sqrt{1-t^2}}{t}\right)+\ln\nu\right)}{2
		\sqrt{1-t^2}}\al,
\\\nn
h_2(t) &=  \frac{1}{16} t^2 \left(5 t^4-6 t^2+1\right)
	-\frac{t^2}{4}  \al,
\\\nn
h_3(t) &=-\frac{t^3 \left(5525 t^6-9945 t^4+4779 t^2-375\right)}{5760}
	+\frac{1}{48}   t \left(15 t^4-8 t^2+\frac{1}{\sqrt{1-t^2}}-1\right)\al
\\\nn&~~~~~~	 + \frac{  t^3}{8(1- t^2)}\al^2.
\\\nn
h_{5}(t) &=\frac{-\al^3}{48}\left(\frac{1}{t^2}-1\right)^{-3/2}.
}
Some of the new coefficients are polynomials in $t$ and in this sense repeat the Bessel case; but not all. The $\al$-term for $k=1$ and the $\al^2$-term for $k=3$ are non-polynomial. For $k=3$, the term with $\al$   is polynomial, but not contained in the Bessel part. The contribution from $k=5$ is new. The polynomial expressions can be calculated by methods similar to those used earlier. First one integrates over $t$ using (3.21) in \cite{bord96-37-895}, then one applies the Abel-Plana formula, for instance (A.1) and (A.2) in \cite{eliz98-31-1743} or (D.5) in \cite{kirs01b}. Note that the cases $k=-1$ and $k=0$ need a special treatment as shown in the mentioned papers.

We split the coefficients into those which are polynomial in $t$ and call them 'Bessel' like coefficients, and the non-polynomial ones, which we  call 'Kummer' coefficients,
\eq{A1.8}{ \ln f_k(t) &= h^B_{p,k}(t) +h^K_{p,k}(t),\qquad (k=1,2,3).
}
In the Kummer coefficients we return to the variable $z$, using \Eq{A1.3}, and get
\eq{A1.9}{ h^K_{1,1}(z) &=
	 \frac{1}{2z}\left[\ln\left(z+\sqrt{1+z^2}\right)+\ln\nu\right],
	 & h^K_{1,3}(z) &= \frac{1}{48z}
\\\nn	h_{1,2}^K(z) &=
\frac{-1}{48\sqrt{1+z^2}}-\frac{1}{8(1+z^2)^{3/2}}
+\frac{5}{16(1+z^2)^{5/2}} +\frac{1}{48z},\  &h^K_{2,3}(z) &= \frac{\sqrt{1+z^2}}{8z},
\\\nn  h^K_{3,5}(z) &= -\frac{1}{48z^3}.
}

\section{\label{B}The 'Temme' coefficients}

Here we display the coefficients of the asymptotic expansion as given in \cite{temm22-33-1221}, and use the notations introduced there. The first coefficient is given in this paper, all higher ones can be calculated recursively. The next two were provided in a private communication by N.Temme.
\begin{verbatim}
	fntilde[1]=
	-1/12*(mu^2*tau^5 - 13*mu^2*tau^4 - mu*tau^4 + 21*mu*tau^3 +
	4*mu*tau^2 - 9*tau^2 - 2*tau - 1)*mu*
	tau^2/(mu*tau^2 - 1)^3/(mu*tau - 1);
\end{verbatim}
\begin{verbatim}
	fntilde[2]=
	1/288*mu*
	tau^4*(mu^5*tau^10 - 26*mu^5*tau^9 + 313*mu^5*tau^8 -
	2*mu^4*tau^9 + 68*mu^4*tau^8 - 1690*mu^4*tau^7 + mu^3*tau^8 +
	1048*mu^4*tau^6 - 60*mu^3*tau^7 + 3255*mu^3*tau^6 -
	3958*mu^3*tau^5 + 18*mu^2*tau^6 + 186*mu^3*tau^4 -
	2678*mu^2*tau^5 + 5462*mu^2*tau^4 - 490*mu^2*tau^3 +
	801*mu*tau^4 - 8*mu^2*tau^2 - 3276*mu*tau^3 + 454*mu*tau^2 +
	4*mu*tau + 720*tau^2 + mu -
	144*tau)/(mu*tau - 1)^2/(mu*tau^2 - 1)^6;
\end{verbatim}

\begin{verbatim}
	fntilde[3]=
	1/51840*tau^4*mu*(139*mu^8*tau^17 + 195*mu^8*tau^16 -
	4695*mu^8*tau^15 - 417*mu^7*tau^16 + 56201*mu^8*tau^14 -
	2001*mu^7*tau^15 + 30105*mu^7*tau^14 + 417*mu^6*tau^15 -
	753675*mu^7*tau^13 + 4848*mu^6*tau^14 + 829668*mu^7*tau^12 -
	69141*mu^6*tau^13 - 139*mu^5*tau^14 + 3197085*mu^6*tau^12 -
	4473*mu^5*tau^13 - 6330048*mu^6*tau^11 + 73563*mu^5*tau^12 +
	1797159*mu^6*tau^10 - 6394821*mu^5*tau^11 + 1431*mu^4*tau^12 +
	19013532*mu^5*tau^10 - 36663*mu^4*tau^11 - 10070166*mu^5*tau^9 +
	6694593*mu^4*tau^10 + 700264*mu^5*tau^8 - 29304894*mu^4*tau^9 +
	6831*mu^3*tau^10 + 22957125*mu^4*tau^8 - 3562623*mu^3*tau^9 -
	3181371*mu^4*tau^7 + 24696258*mu^3*tau^8 + 18579*mu^4*tau^6 -
	27110118*mu^3*tau^7 + 763101*mu^2*tau^8 + 5732235*mu^3*tau^6 -
	10855458*mu^2*tau^7 - 75291*mu^3*tau^5 + 17316315*mu^2*tau^6 +
	1668*mu^3*tau^4 - 5103788*mu^2*tau^5 + 1951776*mu*tau^6 +
	114675*mu^2*tau^4 - 5572800*mu*tau^5 - 5586*mu^2*tau^3 +
	2244240*mu*tau^4 - 139*mu^2*tau^2 - 81648*mu*tau^3 +
	680400*tau^4 + 6480*mu*tau^2 - 388800*tau^3 + 432*mu*tau +
	21600*tau^2 - 1728*tau - 432)/(mu*tau - 1)^3/(mu*tau^2 - 1)^9;
\end{verbatim}
\section{The Abel-Plana formula}
For summation over half-integers, this well known formula reads ($\nu=l+1/2$)
\eq{A2.1}{ \sum_{l=0}^\infty f(\nu) &=
	\int_0^\infty d\nu \ f(\nu)
	+\int_0^\infty \frac{d\nu}{1+\exp(2\pi\nu)} \,g(\nu),
	\quad\mbox{with}\quad
	g(\nu)=\frac{1}{i}\left(f(i\nu)-f(-i\nu)\right).
}
In the text we will refer to the two integrals as 'part A' and 'part B'.

\section{\label{Ahk}The heat kernel coefficients for the interior of a sphere in 3 dimensions (Bessel case)}
These coefficients were initially derived in \cite{bord95-179-215}, recently reprinted in \cite{bord26-8-79}, eq *(15), and read
\eq{hk}{ a_0&=\frac{4}{3}\pi R^3,& a_{1/2}&=-2\pi^{3/2}R^2; &a_1&=\frac{8}{3}\pi  R,
	& a_{3/2}&=-\frac{1}{6}\pi^{3/2},& a_2&=-\frac{16}{315}\frac{\pi}{R}.
}

\section{\label{E}Display of some long formulas}
In this appendix we collect a number of very long formulas which were inconvenient to show in the main text.

In eq. \Eq{2.5.15} we use
\eq{E.1}{
		\tilde p& =
		3\alpha^3 e^{\frac{t(\alpha+z)}{2z}}\left(e^{2t}-1\right)\gamma\, t^3 \\\nn
		&\quad + 3\alpha^2 e^{\frac{t(\alpha+z)}{2z}} t^2\left[\left(e^{2t}-1\right)t
		+ \gamma\left(4+4e^{2t}(t-1)-2t+6e^t t\right)\right]z \\\nn
		&\quad - \alpha t\Bigg[-4e^{\frac{t(\alpha+z)}{2z}}\left(e^{2t}-1\right)(t-3)t \\\nn
		&\qquad\qquad + 3\gamma\Bigg(-8+8e^{2t}
		-12e^{\frac{t(\alpha+3z)}{2z}}(t-2)t \\\nn
		&\qquad\qquad\qquad\qquad + e^{\frac{t(\alpha+5z)}{2z}}\left(-8+12t-5t^2\right)
		+ e^{\frac{t(\alpha+z)}{2z}}\left(8-4t+t^2\right)\Bigg)\Bigg]z^2 \\\nn
		&\quad -\Bigg[48+48e^t(1+3\gamma)t
		+48e^{2t}(t-1+\gamma t) \\\nn
		&\qquad\qquad + e^{\frac{t(\alpha+z)}{2z}}t\left(24-8t+t^2\right)
		-18e^{\frac{t(\alpha+3z)}{2z}}\gamma t\left(8-4t+t^2\right) \\\nn
		&\qquad\qquad - e^{\frac{t(\alpha+5z)}{2z}}t\left(24-8t+t^2
		+6\gamma\left(8-4t+t^2\right)\right)\Bigg]z^3 \\\nn
		&\quad -3t\Bigg[-\alpha^3 e^{\frac{t(\alpha+z)}{2z}}\left(e^{2t}-1\right)t^2
		+ 2\alpha^2 e^{\frac{t(\alpha+z)}{2z}} t\left(-2-2e^{2t}(t-1)+t-3e^t t\right)z \\\nn
		&\qquad\qquad + \alpha\Bigg(-8+8e^{2t}
		-12e^{\frac{t(\alpha+3z)}{2z}}(t-2)t
		+ e^{\frac{t(\alpha+5z)}{2z}}\left(-8+12t-5t^2\right) \\\nn
		&\qquad\qquad + e^{\frac{t(\alpha+z)}{2z}}\left(8-4t+t^2\right)\Bigg)z^2
		- 2e^t\left(3+e^t\right)\left(-8+e^{\frac{t(\alpha+z)}{2z}}\left(8-4t+t^2\right)\right)z^3\Bigg]\ln t.
}
In eq. \Eq{2.5.16} we defined
\eq{E.2}{
q(z) &= \frac{1}{24z^2}\Bigg(-9\alpha^3\pi^2 - 12\alpha^2 z - 12\alpha^2\gamma z\
- 33\alpha^2\pi^2 z - 24\alpha\gamma z^2 \\\nn
&\qquad\qquad - 35\alpha\pi^2 z^2 + 8z^3 - 12\gamma z^3 - 11\pi^2 z^3
- 8\alpha^3\pi^2\ln 2 \\\nn
&\qquad\qquad - 24\alpha^2 z\ln 2 - 24\alpha^2\pi^2 z\ln 2
- 24\alpha\pi^2 z^2\ln 2 + 12z^3\ln 2 \\\nn
&\qquad\qquad - 8\pi^2 z^3\ln 2 - 48z^3\ln A
- 36\alpha^3\zeta'(2) - 108\alpha^2 z\,\zeta'(2) \\\nn
&\qquad\qquad - 108\alpha z^2\zeta'(2) - 36z^3\zeta'(2)\Bigg) \\\nn
&\quad + \frac{1}{8z^2}\Bigg(-4\alpha^2\gamma z
+ 8\gamma z^3\,\zeta\!\left(-2,1+\tfrac{\alpha}{2z}\right)
- 8\gamma z^3\,\zeta\!\left(-2,3+\tfrac{\alpha}{2z}\right) \\\nn
&\qquad\qquad - 8\alpha\gamma z^2\,\zeta\!\left(-1,1+\tfrac{\alpha}{2z}\right)
+ 8z^3\,\zeta\!\left(-1,1+\tfrac{\alpha}{2z}\right)
+ 8\gamma z^3\,\zeta\!\left(-1,1+\tfrac{\alpha}{2z}\right) \\\nn
&\qquad\qquad + 8z^3\,\zeta\!\left(-1,2+\tfrac{\alpha}{2z}\right)
+ 8\alpha\gamma z^2\,\zeta\!\left(-1,3+\tfrac{\alpha}{2z}\right)
+ 24\gamma z^3\,\zeta\!\left(-1,3+\tfrac{\alpha}{2z}\right) \\\nn
&\qquad\qquad + 2\alpha^2\gamma z\,\zeta\!\left(0,1+\tfrac{\alpha}{2z}\right)
- 8\alpha z^2\,\zeta\!\left(0,1+\tfrac{\alpha}{2z}\right)
- 4\alpha\gamma z^2\,\zeta\!\left(0,1+\tfrac{\alpha}{2z}\right) \\\nn
&\qquad\qquad + 8z^3\,\zeta\!\left(0,1+\tfrac{\alpha}{2z}\right)
- 8\alpha z^2\,\zeta\!\left(0,2+\tfrac{\alpha}{2z}\right)
- 8z^3\,\zeta\!\left(0,2+\tfrac{\alpha}{2z}\right) \\\nn
&\qquad\qquad - 2\alpha^2\gamma z\,\zeta\!\left(0,3+\tfrac{\alpha}{2z}\right)
- 12\alpha\gamma z^2\,\zeta\!\left(0,3+\tfrac{\alpha}{2z}\right)
- 16\gamma z^3\,\zeta\!\left(0,3+\tfrac{\alpha}{2z}\right) \\\nn
&\qquad\qquad - 2\alpha^2 z\,\psi\!\left(0,1+\tfrac{\alpha}{2z}\right)
+ 4\alpha z^2\,\psi\!\left(0,1+\tfrac{\alpha}{2z}\right)
- 2\alpha^2 z\,\psi\!\left(0,2+\tfrac{\alpha}{2z}\right) \\\nn
&\qquad\qquad - 4\alpha z^2\,\psi\!\left(0,2+\tfrac{\alpha}{2z}\right)
- \alpha^3\,\gamma_1\!\left(1+\tfrac{\alpha}{2z}\right)
+ 4\alpha z^2\,\gamma_1\!\left(1+\tfrac{\alpha}{2z}\right) \\\nn
&\qquad\qquad - 6\alpha^2 z\,\gamma_1\!\left(2+\tfrac{\alpha}{2z}\right)
- 12\alpha z^2\,\gamma_1\!\left(2+\tfrac{\alpha}{2z}\right)
+ \alpha^3\,\gamma_1\!\left(3+\tfrac{\alpha}{2z}\right) \\\nn
&\qquad\qquad + 6\alpha^2 z\,\gamma_1\!\left(3+\tfrac{\alpha}{2z}\right)
+ 8\alpha z^2\,\gamma_1\!\left(3+\tfrac{\alpha}{2z}\right)
- 8z^3\,\zeta^{(1,0)}\!\left(-2,1+\tfrac{\alpha}{2z}\right) \\\nn
&\qquad\qquad + 8z^3\,\zeta^{(1,0)}\!\left(-2,3+\tfrac{\alpha}{2z}\right)
+ 12\alpha z^2\,\zeta^{(1,0)}\!\left(-1,1+\tfrac{\alpha}{2z}\right)
+ 24z^3\,\zeta^{(1,0)}\!\left(-1,2+\tfrac{\alpha}{2z}\right) \\\nn
&\qquad\qquad - 12\alpha z^2\,\zeta^{(1,0)}\!\left(-1,3+\tfrac{\alpha}{2z}\right)
- 24z^3\,\zeta^{(1,0)}\!\left(-1,3+\tfrac{\alpha}{2z}\right)
- 6\alpha^2 z\,\zeta^{(1,0)}\!\left(0,1+\tfrac{\alpha}{2z}\right) \\\nn
&\qquad\qquad + 8z^3\,\zeta^{(1,0)}\!\left(0,1+\tfrac{\alpha}{2z}\right)
- 24\alpha z^2\,\zeta^{(1,0)}\!\left(0,2+\tfrac{\alpha}{2z}\right)
- 24z^3\,\zeta^{(1,0)}\!\left(0,2+\tfrac{\alpha}{2z}\right) \\\nn
&\qquad\qquad + 6\alpha^2 z\,\zeta^{(1,0)}\!\left(0,3+\tfrac{\alpha}{2z}\right)
+ 24\alpha z^2\,\zeta^{(1,0)}\!\left(0,3+\tfrac{\alpha}{2z}\right)
+ 16z^3\,\zeta^{(1,0)}\!\left(0,3+\tfrac{\alpha}{2z}\right)\Bigg)
,}
where $\zeta(s,a)$ is a Hurwitz zeta function and $\gamma_1(x)$ is the first Stieltjes constant.

\end{document}